\documentclass[11pt]{article}
\usepackage[normalem]{ulem}
\pdfoutput=1 

\usepackage{jheppub} 
\usepackage{url}
\usepackage{caption}
\usepackage{subcaption}
\usepackage{extarrows}

\usepackage[latin9]{inputenc}
\usepackage{float}
\usepackage{amsmath}
\usepackage{amssymb}
\usepackage{graphicx}
\usepackage{esint}
\usepackage{hyperref}
\usepackage{comment}
\usepackage{color}
\usepackage{microtype}
\usepackage{cleveref}
\usepackage{breakurl}
\usepackage{bbm}
\newcommand{\be}{\begin{equation}}
\newcommand{\ee}{\end{equation}}
\newcommand{\ben}{\begin{displaymath}}
\newcommand{\een}{\end{displaymath}}
\newcommand{\bea}{\begin{eqnarray}}
\newcommand{\eea}{\end{eqnarray}}
\def\K{K{\"a}hler }
   \newcommand{\rf}[1]{(\ref{#1})}

\def\be{\begin{equation}}
\def\ee{\end{equation}}
\def\bea{\begin{eqnarray}}
\def\eea{\end{eqnarray}}
\def\ba{\begin{array}}
\def\ea{\end{array}}
\def\bit{\begin{itemize}}
\def\eit{\end{itemize}}

\newcommand{\N}{\mathcal{N}}
\newcommand{\cN}{\mathcal{N}}

 \makeatletter

\allowdisplaybreaks

\makeatother

\makeatletter
\DeclareRobustCommand{\rcite}[1]{%
  \rcite@aux#1,\@nil{#1}%
}
\def\rcite@aux#1,#2\@nil#3{%
  \if\relax#2\relax
    Ref.~\cite{#3}%
  \else
    Refs.~\cite{#3}%
  \fi
}
\makeatother

\hypersetup{
    colorlinks = true,
    citecolor = {blue},
    linkcolor = {blue},
    urlcolor = {blue},
}

 \title{\rm { \LARGE \bf   IIB String Theory and  Sequestered  Inflation  }}

\author[a]{Renata Kallosh,}
\author[a]{Andrei Linde,}
\author[b]{Timm Wrase,}
\author[c]{and Yusuke Yamada}

\affiliation[a]{Stanford Institute for Theoretical Physics and Department of Physics,\\ Stanford University, Stanford, CA 94305, USA}
\affiliation[b]{Department of Physics, Lehigh University, 16 Memorial Drive East, Bethlehem, PA 18018, USA
}
\affiliation[c]{Research Center for the Early Universe (RESCEU), Graduate School of Science,\\ The University of Tokyo, Hongo 7-3-1
Bunkyo-ku, Tokyo 113-0033, Japan}
\emailAdd{kallosh@stanford.edu}
\emailAdd{alinde@stanford.edu}
\emailAdd{timm.wrase@lehigh.edu}
\emailAdd{yamada@resceu.s.u-tokyo.ac.jp}
\notoc
\preprint{RESCEU-15/21}

\abstract{We develop sequestered inflation models, where inflation occurs along flat directions in supergravity models derived from  type IIB string theory. It is compactified on a ${\mathbb{T}^6\over \mathbb{Z}_2 \times \mathbb{Z}_2}$ orientifold with generalized fluxes and O3/O7-planes.  At Step I, we use flux potentials which 1) satisfy tadpole cancellation conditions and 2) have supersymmetric Minkowski vacua with flat direction(s). The 7 moduli are split into heavy  and massless Goldstone multiplets. At  Step II we add a nilpotent multiplet and  uplift the flat direction(s) of the type IIB string theory to phenomenological inflationary plateau potentials: $\alpha$-attractors with 7 discrete values $3\alpha = 1, 2, 3, ..., 7$. Their  cosmological predictions are determined  by the hyperbolic geometry  inherited from string theory. The masses of the heavy fields and the volume of the extra dimensions change during inflation, but this does not affect the inflationary dynamics. }

\begin{document}

\maketitle

   \newpage

 \parskip 5pt

\tableofcontents{}

  \parskip 9pt

  \newpage

\section{Introduction}

Our main goal in this paper  is to construct the  models  where inflation can peacefully coexist  with  steep string theory potentials.  To achieve this goal, we will try to find string theory potentials with supersymmetric Minkowski flat directions, and then gently uplift them. 
Models of this type were  introduced in the context of M-theory compactified on twisted 7-tori with $G_2$-holonomy  \cite{Gunaydin:2020ric}.  The general structure of sequestered inflation is explained  in our recent paper \cite{Kallosh:2021fvz}, which also contains several simple examples illustrating our scenario. In this paper we will apply these methods to IIB string theory. 

The method consists of two steps. At Step I, we will try to find supersymmetric Minkowski vacua with flat directions originating from IIB string theory.  The goal is to find either a single flat direction looking like a  bottom of a mountain gorge, or several different valleys separated from each other by extremely high barriers.  At Step II, we will introduce a nilpotent field  and uplift these flat directions, transforming them into plateau inflationary potentials. The height of these plateau potentials can be many orders of magnitude smaller than the height of the barriers stabilizing the flat directions.  We find that under certain conditions specified in  \cite{Kallosh:2021fvz}, the superheavy fields involved in the stabilization of the Minkowski vacua in string theory do not interfere with inflation. 

The choice of inflationary potentials at Step II is  phenomenological, we do not derive them from string theory. However, as we will see, the resulting inflationary models belong to the general class of $\alpha$-attractors   \cite{Kallosh:2013hoa,Ferrara:2013rsa,Kallosh:2013yoa,Galante:2014ifa}. An important property of $\alpha$-attractors is stability of their  predictions with respect to the choice of inflationary potentials. Most important observational consequences of these models are determined not by their potentials, but by the hyperbolic geometry of the moduli space  inherited from  string theory.

\begin{figure}[!h]
\begin{center}
\vspace{-1mm}
\hspace{-3mm}
 \includegraphics[scale=0.313]{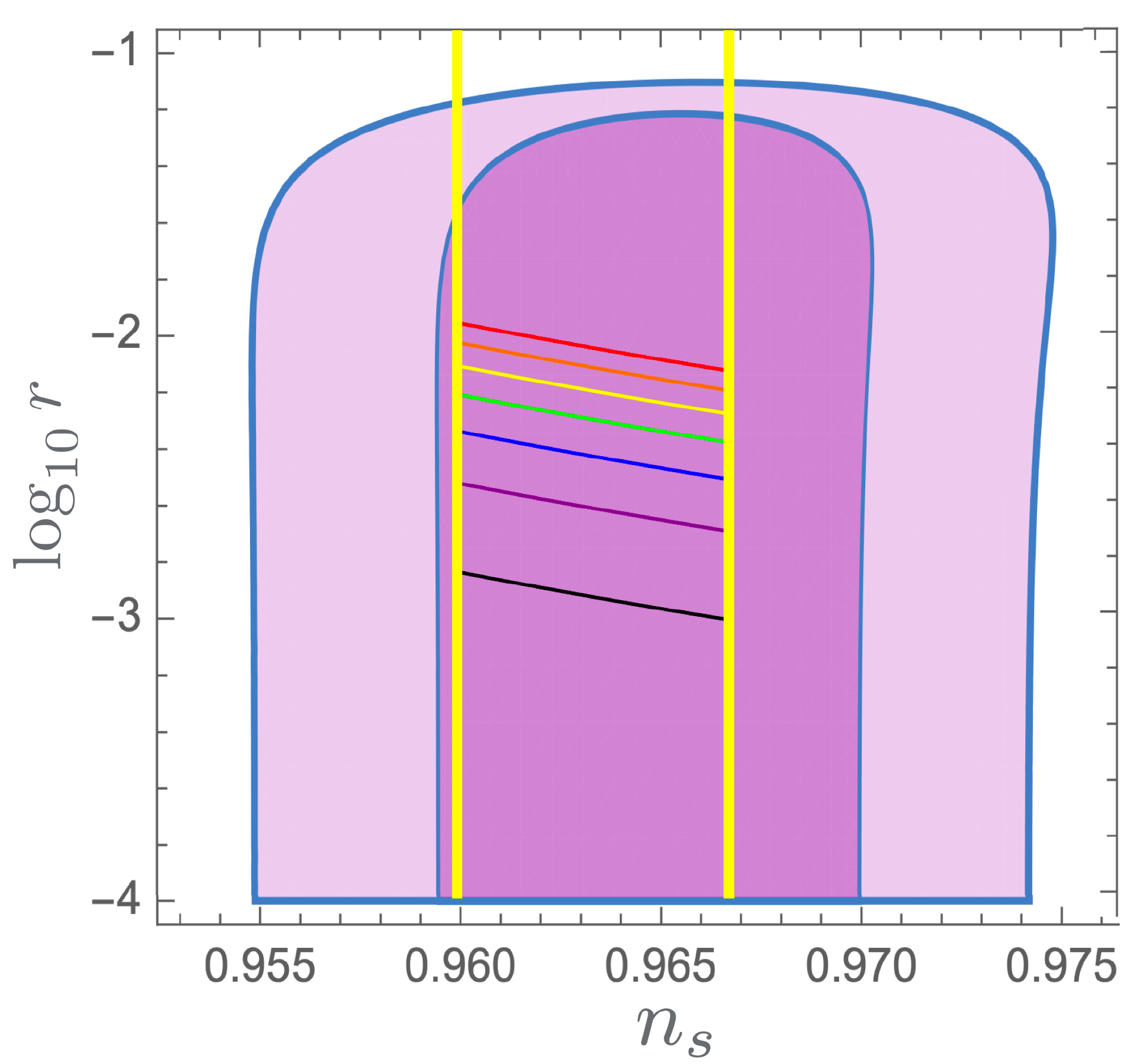}  \hskip 40pt
\includegraphics[scale=0.315]{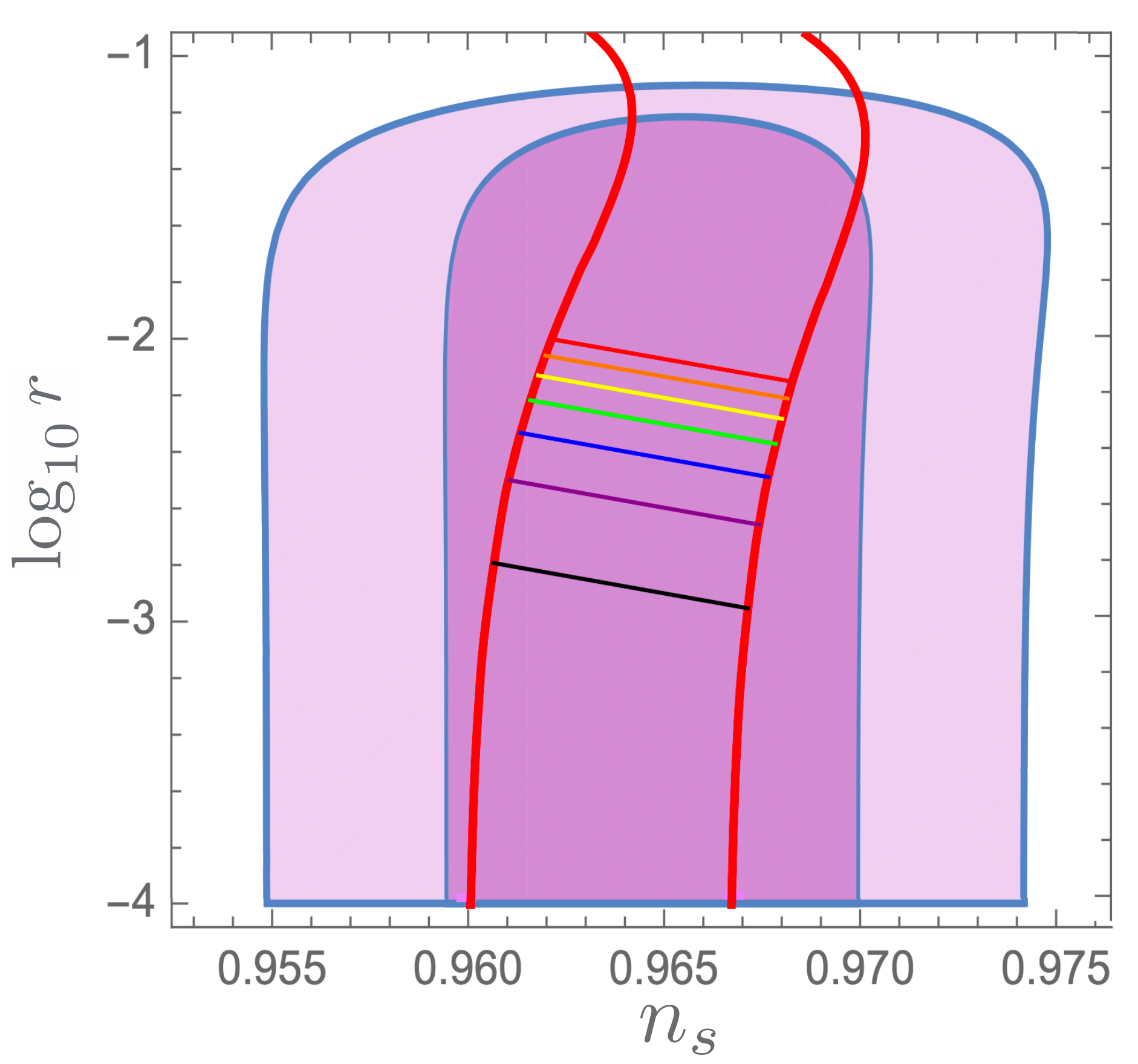}
\end{center}
\vspace{-.4cm}
\caption{\small
  $\alpha$-attractor benchmarks for T-models (left panel) and E-models (right panel).  The predictions are shown for the number of e-foldings in the range $50 < N_e< 60$. Dark pink area corresponds to $n_{s}$ and $r$ favored by Planck2018 after taking into account all CMB-related data. }
\label{7disk2}
\end{figure}

The main predictions of these models matching the observational data are the spectral index $n_{s}$ and the tensor to scalar ratio $r$ for a given number of e-foldings $N_e$: 
\be
 n_{s}= 1-{2\over N_e}, \qquad r = {12 \alpha \over N_e^{2}} \ .
  \ee
These predictions are shown in Fig.~\ref{7disk2} for two  classes of $\alpha$ attractors, T-models, with the inflaton potential $V = V_{0} \tanh^{2}({\phi/ \sqrt{6\alpha}})$, and E-models, with the inflaton potential $V = V_{0} \big (1-e^{-{\sqrt{2/3 \alpha}\,\phi}}\big )^{2}$  \cite{Kallosh:2019hzo}.  We explain the relation between these models  in section \ref{ET}.

In  $\N = 1$ supergravity, parameter $\alpha$ can take any value. The corresponding predictions are shown by the area bounded by two thick yellow lines    for T-models, and by two thick red lines for E-models. However, in some supergravity models originating from M-theory or string theory, the parameter  $3\alpha$ is expected to  take one of the 7 integer values, $3\alpha = 1,2,3,4,5,6,7$~\cite{Ferrara:2016fwe,Kallosh:2017ced,Kallosh:2017wnt}, with predictions shown by 7 parallel lines in each of the two panels in Fig.~\ref{7disk2}. The upper line corresponds to $3\alpha =  7$ with  $r \sim 10^{{-2}}$. It will be the first one of this family of models to be tested by cosmological observations.

The discrete B-mode targets  in the left panel of Fig.~\ref{7disk2} are related to Poincar\'e disks with $V = V_{0} Z\bar Z=V_{0} \tanh^{2}({\phi/ \sqrt{6\alpha}})$ potential. They originate from the \K\, potential defining Poincar\'e disk geometry
\be
K=- 3\alpha \log(1-Z \overline Z) \quad \Rightarrow   \quad ds^2 = 3\alpha {dZ d\overline Z\over (1-Z\overline Z)^2}
\ee
The unit size Poincar\'e disk has $3\alpha=1$ and it is a first from the bottom line in Fig.~\ref{7disk2} at the left panel.

In this paper we will describe the origin of these models and of their predictions in the context of sequestered inflation.
At Step I we will look  for  4D Minkowski vacua  with  one or more flat directions in type IIB string theory compactified on a ${\mathbb{T}^6\over \mathbb{Z}_2 \times \mathbb{Z}_2}$ orientifold with generalized fluxes and O3/O7-planes \cite{Shelton:2005cf,Aldazabal:2006up}. 
At Step II we will uplift the flat directions and derive all seven  cosmological models with $3\alpha = 1,2,3,4,5,6,7$.
We consider type IIB string theory setups having seven moduli chiral superfields $(S, T_I, U_I)$ where $I=1,2,3$. 

Each of the seven moduli has the hyperbolic geometry as its target space geometry,\footnote{A target space hyperbolic geometry here is defined by a \K potential $K= -3\alpha \ln (T+\bar T)$. The case $K= - \ln (T+\bar T)$ in \rf{K} means $3\alpha=1$.}  which can realize $\alpha$-attractors  with $3\alpha=1$,
\begin{eqnarray}
K &= & -\sum_{I=1}^{3}\ln  (T_{I}+\overline{T}_{I})  -\ln (S+\overline{S}) -\sum_{I=1}^{3}\ln (U_{I}+\overline{U}_{I})  \, .
\label{K}\end{eqnarray}
Our aim of Step I is to find string theoretically motivated models where we may stabilize some of the moduli, while keeping some of the inflaton candidates massless. We found two classes of models: The first one has after Step I a single massless superfield, whose moduli space geometry realizes  $3\alpha=7$. At Step II we uplift it to the top Poincar\'e disk target in Fig. \ref{7disk2}. The second model has three unfixed moduli at Step I. They have the geometries of $3\alpha=4, 2, 1$, respectively. At Step II this model will be uplifted to produce all Poincar\'e disk targets in Fig.~\ref{7disk2} above.

The flux superpotential which we use for Step I satisfies all tadpole conditions and Bianchi identities that arise in the presence of our generalized fluxes, see also Appendix \ref{app:Tadpoles} for details. We do not require any exotic sources and we will show the number of D3 and D7 branes necessary for each  flux superpotential.

An important property of flux superpotentials which we use at Step I is that they have certain symmetries which predict the number of  Goldstone supermultiplets (number of flat directions). We present the corresponding Goldstone theorem and the relation to symmetries of the superpotentials in Appendix  \ref{app:Goldstone}.

In Step II we deform supersymmetric Minkowski minima to dS cosmological backgrounds with spontaneously broken supersymmetry, developing the construction proposed in \cite{Kallosh:2017wnt,McDonough:2016der}. Specifically, we add a nilpotent superfield $X$ interacting with the moduli and we add phenomenological superpotential terms, like the mass of the gravitino. This  will lead to  inflation  which can realize the B-mode targets $3\alpha=1,....,7$ in Fig.~\ref{7disk2}. 

The nilpotent superfield $X$ is known to be associated with the anti-D3 brane in string theory \cite{Kachru:2003aw,Ferrara:2014kva,Kallosh:2014wsa,Bergshoeff:2015jxa,Kallosh:2015nia} and more generically with non-supersymmetric branes \cite{Kallosh:2018nrk, Cribiori:2020bgt}. It was proposed in \cite{McDonough:2016der} that the interaction of $X$ with the moduli fields might be a result of the quantum corrections to the \K potential. 
The features of the construction at Step II are to a large degree independent of Step I, as long as a specific choice of a string theory compactification on ${\mathbb{T}^6\over \mathbb{Z}_2 \times \mathbb{Z}_2}$ with fluxes and O3/O7-planes is made and a supersymmetric Minkowski vacuum with a certain number of flat directions is derived. The reason for this sequestering is the fact that $W_{\rm flux} = \partial W_{\rm flux}=0$. Therefore, in the resulting cosmological theory the only trace left of the choice of $W_{\rm flux}$ is how many flat directions there are for different choices. The resulting cosmological models depend only on the choice of the uplifting construction which turns the flat directions of supersymmetric Minkowski vacua into inflationary plateau potentials.

\section{Step I : Type IIB String Theory}
\subsection{Effective 4D theory and tadpole cancellation conditions}
We study an effective 4D $\cN=1$ supergravity theory obtained from type IIB string theory compactified on the orientifold $\mathbb{T}^{6}/\mathbb{Z}_{2}\times \mathbb{Z}_{2}$ with O3/O7-planes.
We define our moduli fields to be $S$,  the 4d complex axion-dilaton,  $U_{I} $ are the complex structure moduli and  $T_{I}$ are the K\"{a}hler moduli.  
The K\"ahler potential is  given in eq. \rf{K}.
The superpotential  also depends on these seven untwisted closed string moduli. It is generated by the presence of 10D NS-NS fluxes $H_3$, R-R fluxes $F_3$ as well as non-geometric $Q$ and $P$ fluxes~\cite{Shelton:2005cf,Aldazabal:2006up}. It involves moduli dependent terms starting from a constant term and up to terms quintic in the moduli. We present the general $W$ for such a compactification in eq.~\rf{general}. There are many tadpole cancellation conditions which such a general flux potential must satisfy, we show them  in eqs.~\rf{D3}-\rf{QP} which are taken from~\cite{Aldazabal:2006up,Guarino:2008ik}.

Here  we will use only terms in $W$ that are quadratic in the fields
\begin{align}
W=-\sum_{I=1}^3 q_I\frac{U_1U_2U_3}{U_I}-\sum_{I=1}^3a_ISU_I-\sum_{I,J=1}^3b_{JI}T_IU_J-\sum_{I=1}^3f_IST_I\,.
\label{quad}\end{align}
As we will see below, this simple form automatically satisfies various tadpole conditions. Nevertheless, there are still several nontrivial conditions, which constrain the parameters in the superpotential.

Below we calculate the tadpole equations in the quotient space, as in \cite{Aldazabal:2006up}. This means we are counting a brane plus its image under the O3/O7 orientifold involution as 1. A brane stuck on an orientifold does not have an image and would therefore be counted as a 1/2 brane. This leads to $N_{D3}, N_{D7}$ being positive half - integers, i.e., we should ensure that for any given model the numbers below are positive integers, half-integers or zero. 

Below we give the list of simplified tadpole conditions for only the fluxes that appear in eq.~\rf{quad}, whereas all tadpole cancellation conditions for the general expression for $W$ are given in appendix \ref{app:Tadpoles}.

\noindent D3 brane number: (2.35) in \cite{Aldazabal:2006up}
\begin{equation}
N_{\rm D3}=16-\frac12 \sum_{I=1}^3q_Ia_I.\label{ND3}
\end{equation}

\noindent D7 brane number: (3.9) in \cite{Aldazabal:2006up} (after adding the contributions from O7-planes, see also equation (3.19) and the text below it in \cite{Guarino:2008ik})
\begin{equation}
N_{{\rm D7}_I}=16-\frac12\sum_{J=1}^3q_Jb_{JI}.\label{ND7}
\end{equation}

\noindent NS7 brane number: (4.13) in \cite{Aldazabal:2006up}
\begin{equation}
N_{{\rm NS7}_I}=0.
\end{equation}

\noindent I7 brane number: (4.40) in \cite{Aldazabal:2006up}
\begin{equation}
N_{{\rm I7}_I}=0.
\end{equation}

\noindent $Q H-PF=0$ constraints: (4.35) in \cite{Aldazabal:2006up} for $I\neq J \neq K \neq I$
\begin{align}
a_Ib_{JJ}+a_Jb_{IJ}-q_Kf_J=0.\label{tad1}
\end{align}

\noindent $QQ=0$ constraints: (3.30) in \cite{Aldazabal:2006up} for $I\neq J \neq K \neq I$
\begin{align}
-b_{II}b_{JK}-b_{JI}b_{IK}=0.\label{tad2}
\end{align}
Note, that we do not need NS7 or I7 branes in our models since their numbers automatically vanish in our setup. The remaining constraints like $PP=0$ and $QP+PQ=0$ are automatically satisfied for our choice of superpotential with only terms that are quadratic in the fields.

\subsection{One flat direction, one Goldstone supermultiplet}\label{one}

\noindent We found two superpotentials leading to a single modulus superfield $T$
\be
T\equiv T_1=T_2= T_3=U_1=U_2=U_3=S.
\label{T}\ee
The low energy effective K\"ahler potential for $T$ becomes
\be
K_7  =- 7\ln (T+\overline{T}).
\label{K7}
\ee
We found two specific superpotentials satisfying the tadpole conditions \eqref{tad1}, \eqref{tad2}. The first one is
\bea
W_7^{(1)} &&= (S - T_1) (U_1 - U_3) + (T_2 - T_3) (U_2 - U_1) + (T_1 -S) (U_3 - U_2)
 \cr
\cr
&& + 2 (U_1 - S) (T_1 - T_3) + 2 (T_2 - T_3) (U_3 - S) \,.\label{eq:W71}
\label{71}\eea
The equations \eqref{ND3} and \eqref{ND7} give the number of D-branes required for this superpotential as
\be
N_{D3}= N_{D7_1} =  N_{D7_2} =  N_{D7_3} = 16
   \label{tad7}
 \ee
to cancel the O3/O7-plane contributions. No other branes are needed and 96 Bianchi identities without exotic sources are satisfied. Note, that the number of D3/D7-branes is exactly the required number to fully cancel the contributions from the O3/O7-planes, i.e., our choice of fluxes that led to the superpotential in eq. \eqref{eq:W71} does not induce any charges. We could therefore also not do the orientifold projection and would not have to include any local sources at all in this particular model.

Another example of a superpotential that leads to the single modulus~\eqref{T} is
\be
W_7^{(2)} =(S-U_1)(T_1-U_2)+(S-U_2)(T_2-U_3)+(S-U_3)(T_3-U_1)\,.\label{72}
\ee
 In this case, the following numbers of branes are required
\be
  N_{D3}= {35\over 2}\,, \qquad 
  N_{D7_1} =     N_{D7_2} =  N_{D7_3} = {33\over 2}\,.
   \label{tad7v2}
\ee
It is interesting  that the superpotentials~\eqref{eq:W71} and \eqref{72} are manifestly invariant under the uniform shift
\begin{equation}\label{eq:shiftsym}
S\to S+C(S, T_I, U_I),\quad T_I\to T_I+C(S, T_I, U_I), \quad U_I\to U_I+C(S, T_I, U_I),
\end{equation}
where $C(S, T_I, U_I)$ is a   holomorphic function of all moduli.  
This is a symmetry predicting one flat direction in this model, i.e., one Goldstone supermultiplet, as explained in Appendix \ref{app:Goldstone}.

\subsection{Three flat directions, three Goldstone supermultiplets}\label{three}
 \noindent We also found another class of models where we are left with three moduli superfields. We call this class of models  {\it split} (4,2,1) disk models. It  is realized by the superpotential 
 \bea
W_{(4,2,1)} =(U_1 - U_3) (T_1 - T_3) + (S - U_2) (T_3 - U_1+T_1 - U_3)\,.
\label{split2}\eea
The scalar potential has a Minkowski minimum with three flat directions:
\be
T_{(1)} \equiv T_1= T_3= U_1=U_3,  \qquad T_{(2)} \equiv S=U_2, \qquad T_{(3)} \equiv T_2 \,.
\label{3flat}\ee
After integrating out the heavy modes, the \K potential in eq. \rf{K} becomes
\be
K_{(4,2,1)}  = - 4 \ln (T_{(1)}+\overline{T}_{(1)}) - 2 \ln (T_{(2)}+\overline{T}_{(2)})  -  \ln (T_{(3)}+\overline{T}_{(3)}) \,.
\label{K3flat}\ee
The above superpotential satisfies the tadpole conditions~\eqref{tad1} and \eqref{tad2}, and from \eqref{ND3} and \eqref{ND7} we find that the following branes are required
\be
N_{D3}=17\, , \qquad N_{D7_1} =  N_{D7_2} =  N_{D7_3} = 16\,.
 \label{tad3flat}\ee
Let us briefly look at the symmetry structure of this model. The superpotential is invariant under the following three shift symmetries
\bea
T_{(1)}(=U_1,U_3,T_1,T_3) &\to& T_{(1)} + C_1(S, T_I, U_I)\,,\cr
T_{(2)}(=S,U_2) &\to& T_{(2)}+C_2(S, T_I, U_I)\,,\cr
T_{(3)}(=T_2) &\to& T_{(3)}+C_3(S, T_I, U_I)\,, 
\label{sym}\eea
where $C_{1,2,3}(S, T_I, U_I)$ are  holomorphic functions of all moduli. As was the case for the one flat direction models, the symmetry generators of the superpotential are in one-to-one correspondence with the massless moduli multiplets, namely, there are three Goldstone supermultiplets here in agreement with the Goldstone theorem in Appendix \ref{app:Goldstone}.

Thus, it turns out that the symmetry of the flux superpotential is responsible for the realization of interesting cosmological models, which are discussed later.

\subsection{Mass eigenvalues: increasing/decreasing moduli (volume) during inflation}
In this subsection we study the values of the masses squared of the heavy fields when the massless field changes.  For that purpose we  calculate how masses change when we move along the flat modulus direction towards either larger or smaller values. We find that the masses decrease when Re\,$T=t$ increases. The masses squared depend exponentially on the canonically normalized inflaton field and could in some case even become tachyonic. However, as we will see in Step II when we introduce an inflationary potential,  ultimately we will be able to have of order hundred e-folds of inflation and heavy masses that stay above the Hubble mass and below a few $M_{\rm pl}$.

Concretely, with $T= \overline{T} = t= e^{-\sqrt{2\over 7} \phi}$ for the case with one flat direction and the model in eq. \rf{71} with the \K potential in eq. \rf{K7} we find the eigenvalues of the squared masses of our seven multiplets to be\footnote{In a simplified model with three moduli in \cite{Kallosh:2021fvz} and one flat direction we had $T= \overline T = e^{-\sqrt{2\over 3} \phi}$ and the squared mass eigenvalues have a factor $(m^2)^i= (m^2)^i_0 \, t= (m^2)^i_0 e^{-\sqrt{2\over 3}\phi}
$. In general for $n$ moduli the formula is $(m^2)^i= (m^2)^i_0 \, t^{-n+4}$.}
\be
(m^2)^i=(m^2)^i_0 \,  t^{-3} = (m^2)^i_0 \,  e^{3 \sqrt{2\over 7}\phi}
\label{massesStepI}\ee
where 
\be
(m^2)^i_0 \approx \{ 6.0727, \quad 4.0024, \quad 0.7589, \quad 0.5, \quad 0.5, \quad 0.1660, \quad 0\}
\ee
in Planck mass units $M_{\rm pl}=1$.
Here $\phi$ is a canonically normalized real scalar field that ultimately will become the inflaton. We see that at $\phi=0$ there are 6 masses of order $M_{\rm pl}^2$ and one mass is zero. At large positive values of $\phi$ the values of the modulus $t$ are smaller than that at $\phi=0$ and at large negative values of $\phi$ the values of the modulus are larger  than at $\phi=0$. 
\begin{figure}[!h]
\vspace{-2mm}
\hspace{-3mm}
\begin{center}
 \includegraphics[scale=0.30]{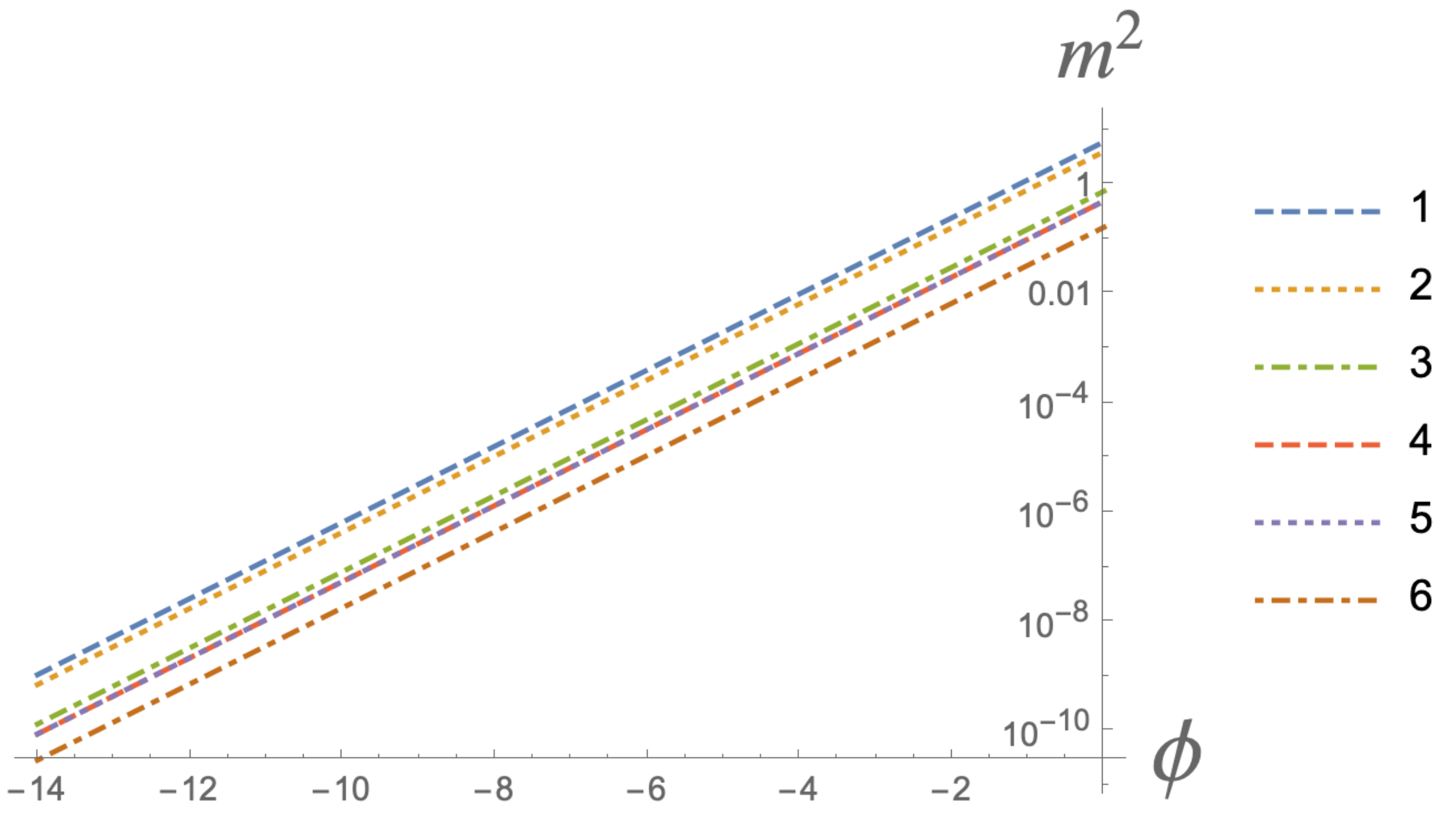}  \hskip 28pt
\includegraphics[scale=0.31]{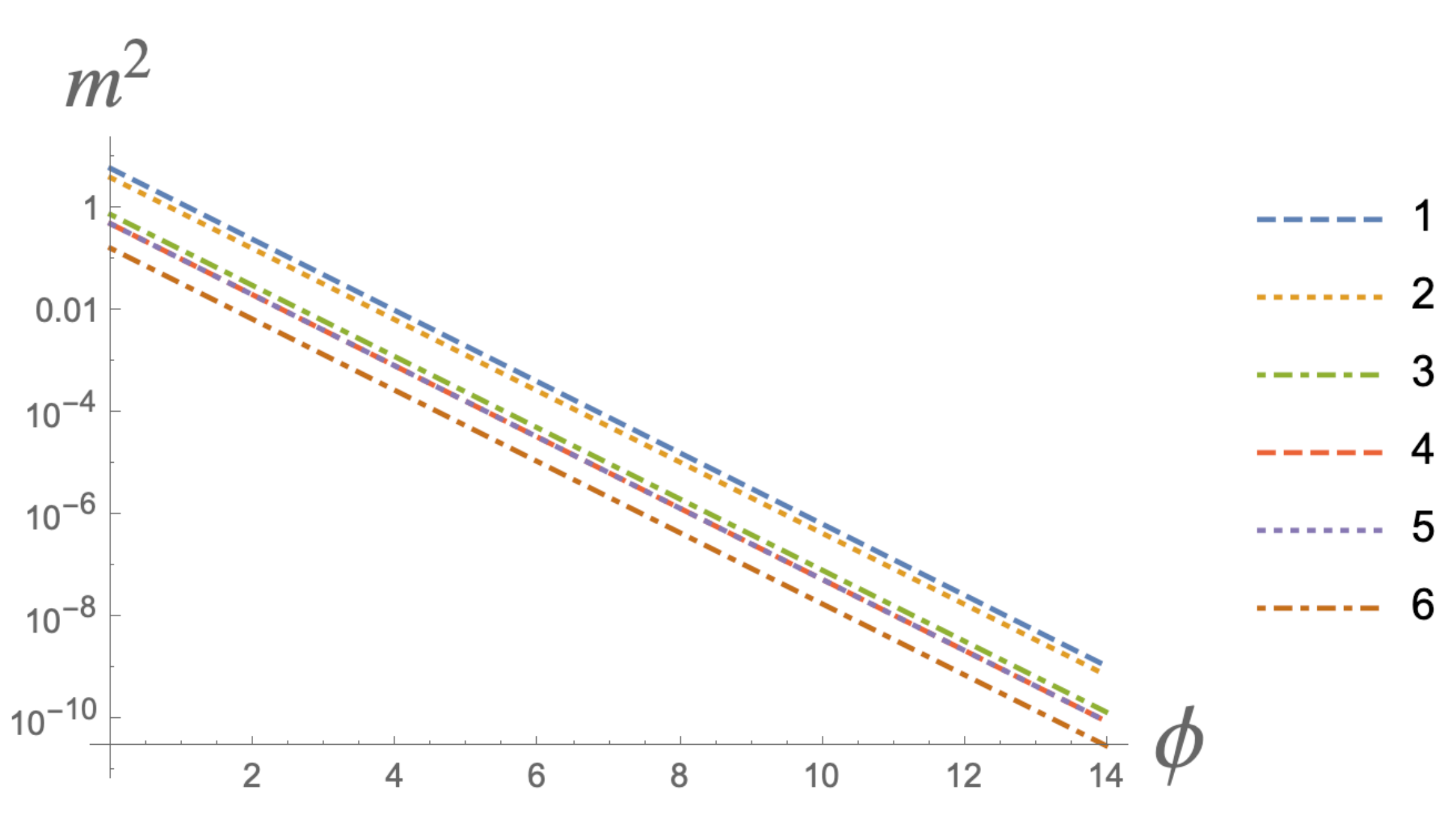}
\end{center}
\vspace{-.2cm}
\caption{\small On the left panel we present a {\it growing volume scenario}: There are six heavy multiplets according to eq. \rf{massesStepI} plotted against the canonical field $\phi$. They start at $\phi=0$ at Planckian scale and decrease to the left, towards the exit from inflation. On the right panel there is a {\it decreasing  volume scenario} : a plot of the six heavy masses with $T = e^{\sqrt{2\over 7} \phi}$.} 
\label{volume}
\end{figure}
\vskip -12pt
This observation suggests that if we would like to start inflation in 4D at small values of the moduli (and small  volume of the extra dimensions) and have them increase during inflation, we might be interested in having the initial stage of inflation at about $\phi=0$ and move towards some negative values of $\phi$. 
The masses of 6 heavy fields decrease with increasing $|\phi|$ and one has to check that at Step II that these heavy fields are not getting too light or even become tachyonic during inflation.

In the opposite case, we can change the canonical field $\phi$ to $-\phi$ and take $T=\overline T = e^{\sqrt{2\over 7} \phi}$. We can start inflation at  large values of the modulus, at some positive value of $\phi$ and move towards $\phi=0$. The masses of the 6 heavy fields will grow during inflation towards smaller $\phi$
\be
(m^2)^i=(m^2)^i_0 \,  t^{-3} = (m^2)^i_0 \,  e^{-3 \sqrt{2\over 7}\phi}
\ee and we have to check that they do not exceed the Planckian scale but still are sufficiently heavy at the beginning of inflation. We have studied these models at Step II and found that they are consistent: there are no tachyons up to more than 500 e-foldings of inflation. Therefore, this scenario with decreasing volume is viable.

We show both of the above discussed scenarios in Fig. \ref{volume}. In the M-theory models in \cite{Gunaydin:2020ric} our choice was the growing volume scenario, but we started there at positive $\phi$ so that at the minimum the value of $T=1$ was reached. Here we will start at $\phi=0$ and move towards some negative $\phi$, so that at the minimum some $T=c$ value can be reached with $c\gg1$.

\section{ Step II: Cosmological Models with B-mode Detection Targets}

Following the proposal for cosmological models in~\cite{Gunaydin:2020ric} and as explained in simple models in~\cite{Kallosh:2021fvz}, we now uplift the flat directions of Step I using additional terms in the action. These additional terms include the gravitino mass parameter $W_0$ and a nilpotent superfield  $\hat X^2=0$ with the Volkov-Akulov \cite{Volkov:1972jx} parameter $F_X$, and an inflationary potential depending on the moduli with the flat directions found in Step I.

In the KKLT construction \cite{Kachru:2003aw,Ferrara:2014kva} the role of the anti-D3-brane associated with the nilpotent superfield is to {\it uplift the $AdS_4$ minimum to a $dS_4$ minimum}. Here we start with a Minkowski flat direction and the role of the nilpotent superfield interacting with the flat direction modulus is to {\it uplift the flat direction to a nearly flat plateau-type inflationary potential}.

\subsection{Step II Potential}
It is convenient here to (re-)label the type IIB fields $T_1, T_2, T_3, U_1, U_2, U_3, S$ as the ones in M-theory which were used in \cite{Gunaydin:2020ric}: $T^1, T^2, T^3, T^4, T^5, T^6, T^7$.
 
We consider the Step II seven disk model \footnote{At $W_0=0$ and $X=0$ the eqs. \rf{EKs} and \rf{EWs} trivially lead to models of Step I.}
\begin{eqnarray}
K &= & -\sum_{i=1}^{7}\ln  (T^{i}+\overline{T}^{i})   + \frac{F_X^2}{F_X^2+V_{\rm infl}(T^i,\overline{T}^i)} X \overline X \, .
\label{EKs}\end{eqnarray}
\be
W=W_{\rm flux}(T^i) +(W_0+ F_X X) {\cal V}^{\frac12}, \qquad {\cal V} =\prod_{i=1}^{7} (2  T^i) \,,
\label{EWs}
\ee
where $W_{\rm flux}$ denotes the flux superpotentials  inherited from IIB string theory and discussed in the previous section and $e^{-K } \Big |_{(\overline T^i  \rightarrow  T^i)} $ defines $\cal {V}$. The holomorphic volume factor $\cal{V}$ is responsible for realizing an approximate shift symmetry for the inflaton, see section 2.1.2 of \cite{Kallosh:2021fvz}. The scalar potential at $X=0$  is\footnote{This is a consequence of the nilpotent condition on a chiral superfield $\hat{X}$, $\hat{X}^2=0$ which makes the complex scalar a fermion bilinear, $X=\frac{1}{\sqrt2 \hat{F}^X}\psi^X\psi^X$ where $\psi^X$ is the Goldstino and $\hat{F}^X$ is the auxiliary scalar component of $\hat{X}$.} 
\begin{equation}
V=e^{K}\left(\sum_i K^{i\bar{\imath}}|D_iW|^2+ (F_X^2+V_{\rm infl}(T^i,\overline{T}^i))|{\cal V}|
-3|W|^2\right)\,.
\end{equation}
At  $ T^i= \overline T^i$ and along the flat directions where $W_{\rm flux}= \partial_i W_{\rm flux}=0$ the total potential of Step II  simplifies dramatically and we get 
\begin{equation}\boxed{
V= \Lambda +V_{\rm infl}(T_{\rm flat}
 ) }\end{equation}
where $\Lambda =F_X^2 -3|W_0|^2>0$.  In our models  flat directions are real  $T_{\rm flat}
=\overline{T}_{\rm flat}$. In the models with one flat direction we have $T_{\rm flat}=T^i$ where $ i=1,\dots, 7$.  In the split model  \rf{split2} $T_{\rm flat}$ has 3 components:  $T_{(1)} \equiv U_1=U_3=T_1=T_3,  \ T_{(2)} \equiv S=U_2,  \ T_{(3)} \equiv T_2$.

At $W_{\rm flux}(T^i) = \partial_k W_{\rm flux}(T^i) =0$ and $ T^i= \overline T^i$, we can identify the mass of the gravitino, $m_{3/2}$, and the auxiliary field of the nilpotent multiplet $\hat{F}^X$ in these models
\begin{equation}
|m_{3/2}|^2 =e^{K}|W|^2=W_0^2\,, \qquad  K_{X\bar{X}}|{\hat{F}}^X|^2 = e^{K}K^{X\bar{X}}|D_XW|^2=F_X^2 + V_{\rm infl} (T_{\rm flat} ).
\end{equation}

For simple choices of $V_{\rm infl}$ here we will derive the E-model version of $\alpha$-attractors \cite{Kallosh:2013hoa,Kallosh:2013yoa}. To derive the T-models we need to change variables as described in the next subsection.

\subsection{From E-models to T-models}\label{ET}
Using  the Cayley transformation we can switch from the half plane variables $T^i$  to the disk variables $Z^i$ as shown in \cite{Carrasco:2015uma}
\begin{equation}
T^i=\frac{1+Z^i}{1-Z^i}.
\label{Caley}\end{equation}
In all $3\alpha$ models we  find that the \K potential, the superpotential and the position of the minimum become functions of disk coordinates.  The flat direction we now call  $Z$. We define the T-models via the change of variables from the E-models given in eq. \rf{Caley}.  However, in the expression for $K_{X\overline X}$ we make a different choice of the function $V_{\rm infl}$, i.e., it is not the one which follows from  a change of variables from the E-models. Thus, we have
 \begin{align}
K=&- \sum_{i=1}^7\log(1-Z^i \overline Z^i)+\frac{F_X^2}{F_X^2+ V_{\rm infl} (Z^i,\overline Z^i)  }X\overline{X},\cr
W=&W_{\rm flux}(T^i (Z^i)) +(W_0+ F_X X) {\cal V}^{1/2} (T^i (Z^i))\,.
\label{top71}\end{align}
The flat directions are now defined in terms of $Z$ variables and at  $ Z^i=\overline{Z}^i$ and at the flat directions where $W_{\rm flux}= \partial_i W_{\rm flux}=0$ the total potential of Step II  is simply
\begin{equation}\boxed{
V= \Lambda +V_{\rm infl}(Z_{\rm flat}
 ) }\end{equation}

\section {One valley cosmological scenario}
\subsection{ E-model of the top B-mode target} \label{emodel}
For the  model with  $3\alpha =7$ the single flat direction $T$ with \K potential $K=-7 \log(T+\overline T)$ at $T=T^i$ where $ i=1,\dots, 7$ was defined in eq. \rf{T} in terms of $T_i$. We make a choice 
\be\label{Echoice}
V_{\rm infl} (T_{\rm flat} ) = m^2 \left( 1-{T\over c} \right) \left(1- {\overline T\over c}\right).
\ee
This choice allows us to have the position of the exit from inflation depending on the choice of the parameter $c$. For $c=1$ the minimum is at $T=1$ and $\phi=0$ and inflation starts at some positive values of $\phi$. 
This is suitable for the class of models with the {\it volume of the extra dimensions decreasing during inflation}, as shown in the right panel of Fig. \ref{volume}. For big positive values of $c$ we can start inflation at about $T=1$ and $\phi=0$ and get down to the minimum at some negative values of $\phi$ and $T= c$. This is suitable for the class of models with the {\it  volume of the extra dimensions increasing during inflation}, as shown in the left panel of Fig. \ref{volume}

It is convenient to switch to canonical variables $\phi$ and $\theta$:
\be
T =  e^{-\sqrt{\frac{2}{7}}\phi}\,\left(1  +{\rm i} \sqrt{\frac{2}{7}}\, \theta\right) \ .
\ee
Here $\phi$ is a canonical inflaton field, and $\theta$ has a canonical normalization in the vicinity of ${\theta} = 0$ which corresponds to the minimum of the potential with respect to $\theta$ during inflation. 
 The total potential of the canonically normalized inflaton field $\phi$  at $\theta = 0$ according to our choice \rf{Echoice}  is
\be\label{Epot}
V^E=\Lambda+  m^2\left(1- {1\over c}e^{-\sqrt{\frac{2}{7}}\phi}\right)^2.
\ee
This is a potential of  the top benchmark target $3\alpha=7$ for B-mode detection \cite{Ferrara:2016fwe,Kallosh:2017ced,Kallosh:2017wnt} in E-models.

We need to verify that the six multiplets which were heavy at Step I are still under control and do not destabilize inflation due to changing masses of the heavy fields. 
We study the case of the moduli and volume increasing during inflation and decreasing masses of the heavy fields with superpotential~\eqref{eq:W71}, as in the left panel in Fig. \ref{volume}. 
\begin{figure}[H]
\centering
\includegraphics[scale=0.32]{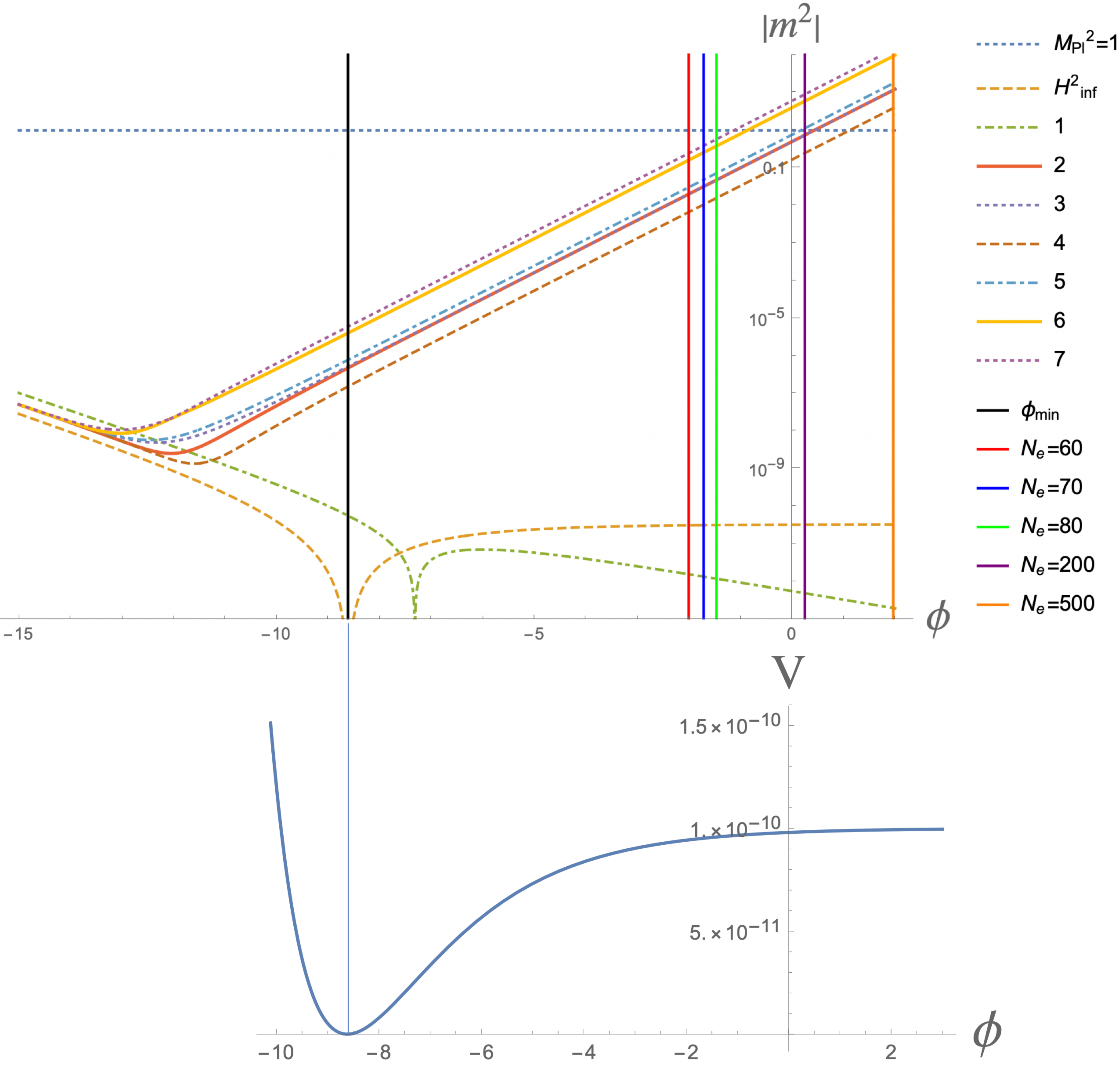}
\caption{\small  At the lower panel we show the inflationary potential as a function of the canonical field $\phi$. At the top panel we plot the log of the absolute values of the masses squared of the six heavy fields. Here we have used superpotential~\eqref{eq:W71}. Each vertical line shows the value of $\phi$ corresponding to the number of e-foldings $N_e$. The dashed green line shows the log of the absolute values of the mass squared of the inflaton field. The kink at $\phi\approx -7.3$ is where the inflaton mass squared changes sign, as we can see in Fig. \ref{Infl}. The Hubble parameter in this figure is approximated as $H_{\rm infl}^2=V/3$ (yellow dashed line), which causes the kink behavior at the minimum of $\phi$.
}\label{top1}
\end{figure}
\begin{figure}[H]
\centering
\includegraphics[scale=0.40]{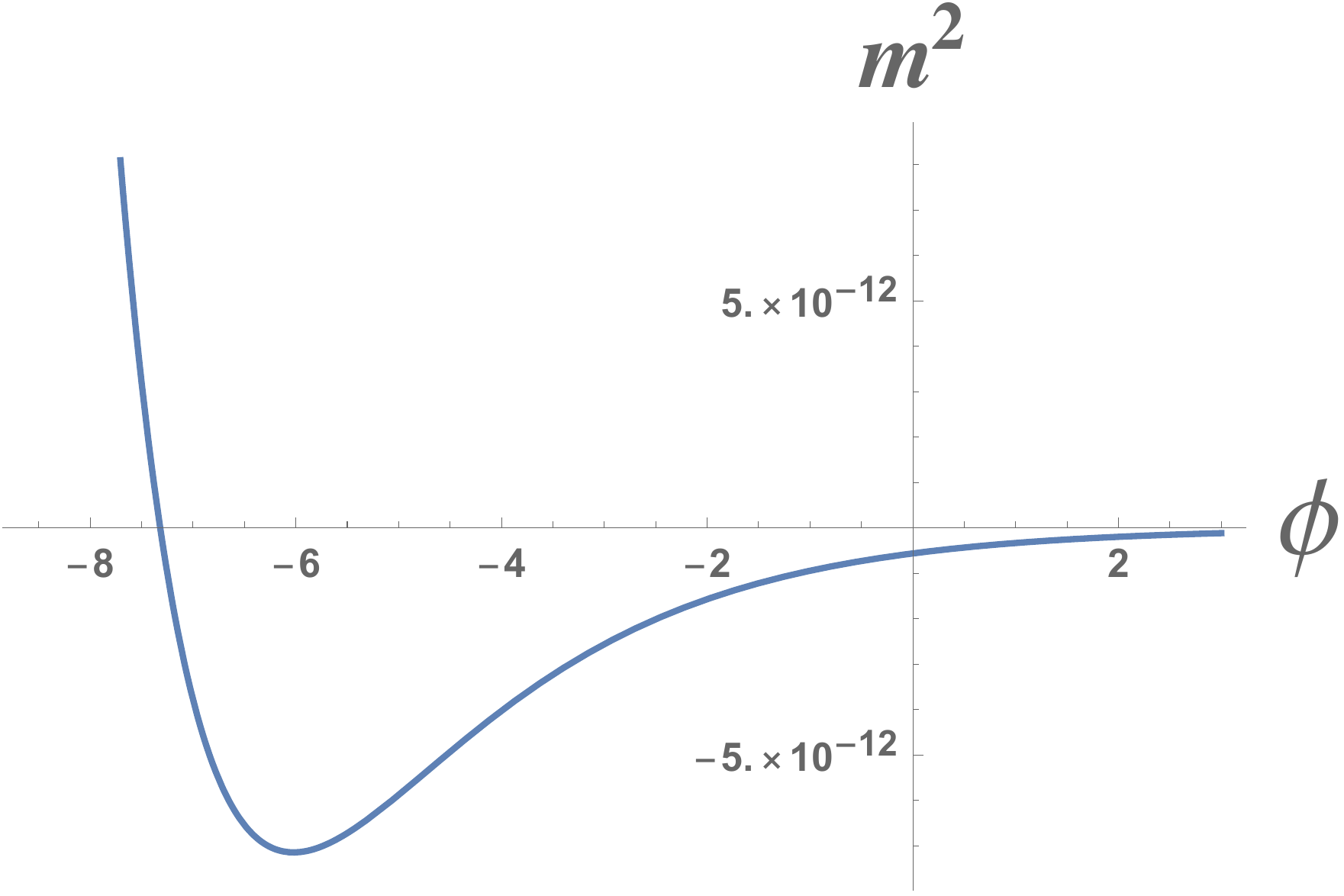}
\caption{\small  Here we plot the mass squared of the inflaton field. It is  negative at the plateau of the potential and flips sign at about $\phi=-7.3$, at which the inflaton mass shows a singular behavior in Fig.~\ref{top1}.}\label{Infl}
\end{figure}
For $c=10^2, W_0= 10^{-5}, m= 10^{-5}, \Lambda =0$ we plot in the Fig. \ref{top1}
the inflationary potential in the lower panel and on the upper panel we plot the log of the absolute values of the masses squared of the six heavy multiplets, as well as  of the inflaton.
The six heavy masses starting from $\phi=0$ to about $\phi=-8$ behave exactly as in Step I in  Fig. \ref{volume} left panel, they decrease from Planckian values down. However, they turn around and go up later at Step II. So, there is no danger of them getting very light or tachyonic.\footnote{This is not true for the axionic masses squared that keep decreasing for more and more negative $\phi$ and they can become tachyonic. However, in the range $-10 \lesssim\phi $ that is relevant for us the axionic masses squared are for our choice of parameters essentially indistinguishable from the corresponding saxion masses squared.}

We note that the kink for the inflaton mass in Fig. \ref{top1} is an artifact of the log of the absolute value of the mass squared, and the kink is a signal of the sign change. To clarify this point we have also plotted the mass  squared of the inflaton in Fig.  \ref{Infl}.
\subsection{Tools for getting lower B-mode targets from split models}
Thus, the model with  $3\alpha =7$ is now described. To get $3\alpha =6,5,4,3,2,1$ we proceed 
with the tools proposed in~\cite{Kallosh:2017ced,Kallosh:2017wnt} for this purpose as well as the ones  used 
in our models in M-theory in~\cite{Gunaydin:2020ric}. There we have found Minkowski vacua with 2 flat directions which we called split models 
\be
(m,n) : \{(6,1), (5,2), (4,3)\}
\label{split}\ee
In these cases we  have a two valley cosmological models with $3\alpha_1 = m$ and $3\alpha_2 = n$.
In  \cite{Gunaydin:2020ric}  the potentials $V_{\rm inf}$ for split models were given  in eq. (7.30). The particular example shown in Fig.~14 there corresponds to the split of the 7 moduli into groups of 6 and 1 discussed above, and the universe is divided into exponentially large parts with $3\alpha_1=6$  and $3\alpha_2=1$, depending on initial conditions. Therefore in some parts we have  B-mode targets with 6 disks in some other with 1 disk. Other split models in \rf{split} will give us all seven B-mode targets.
 
 An additional  set of tools was also suggested in \cite{Kallosh:2017ced,Kallosh:2017wnt} for the purpose of using split models for cosmological purposes. These include fixing  one  of the two flat moduli by just adding to $V_{\rm infl}$ a term that freezes one of the flat moduli (by making it very heavy). The other suggestion was to merge two different disks by adding to $V_{\rm infl}$ an interaction term.
 
In type IIB  models the superpotentials satisfying the tadpole cancellation conditions did not give us the two flat directions models at Step I, however, we have found  three flat direction models, the split $(4,2,1)$ models. We will show below that we have 2 options here. The first is to observe that with three valley cosmological scenario we can have models with $3\alpha = 4,2,1$, depending on initial conditions. However, the cases $3\alpha = 6,5,3$ still have to be obtained differently. For this purpose we will reduce our split $(4,2,1)$ model to the cases studied before, and given in eq. \rf{split}.

\section{Three valley cosmological scenario, $\mathbf{3\alpha = 4,2,1}$}
We study here a model with the superpotential in eq. \rf{split2}, which has three flat $T_{(i)}$ directions defined in eq. \rf{3flat}.
We choose  $V_{\rm infl}$ to be
\be
V_{\rm infl} (T_{(i)} ) = m^2 \sum_{i=1}^3\left(1-{T_{(i)}\over c} \right) \left(1- {\overline T_{(i)}\over c}\right)\,.
\label{Echoice3}\ee
In case of two flat directions this was our choice corresponding to the two valley cosmological scenario in \cite{Gunaydin:2020ric}, however, here we made a shift of the minimum to $T_{(i)}=c$ to study a cosmological scenario with the volume increasing during inflation. At this point, as in the two flat valley, we can already identify the cases with $3\alpha_1 = 4$ and $3\alpha_2 = 2$ and $3\alpha_3 = 1$, based on initial conditions being in vicinity of one of the three valleys.

But we can also use the other tool mentioned above, namely we can enforce via extra terms in $V_{\rm infl} $ two of the light fields to be fixed and only allow one direction in this three dimensional moduli space to serve as the inflaton. In such case, only one field remains as a dynamical inflaton, the other fields are either heavy and changing during inflation or fixed.

\begin{figure}[!h]
\vspace{-1mm}
\hspace{-3mm}
\begin{center}
 \includegraphics[scale=0.23]{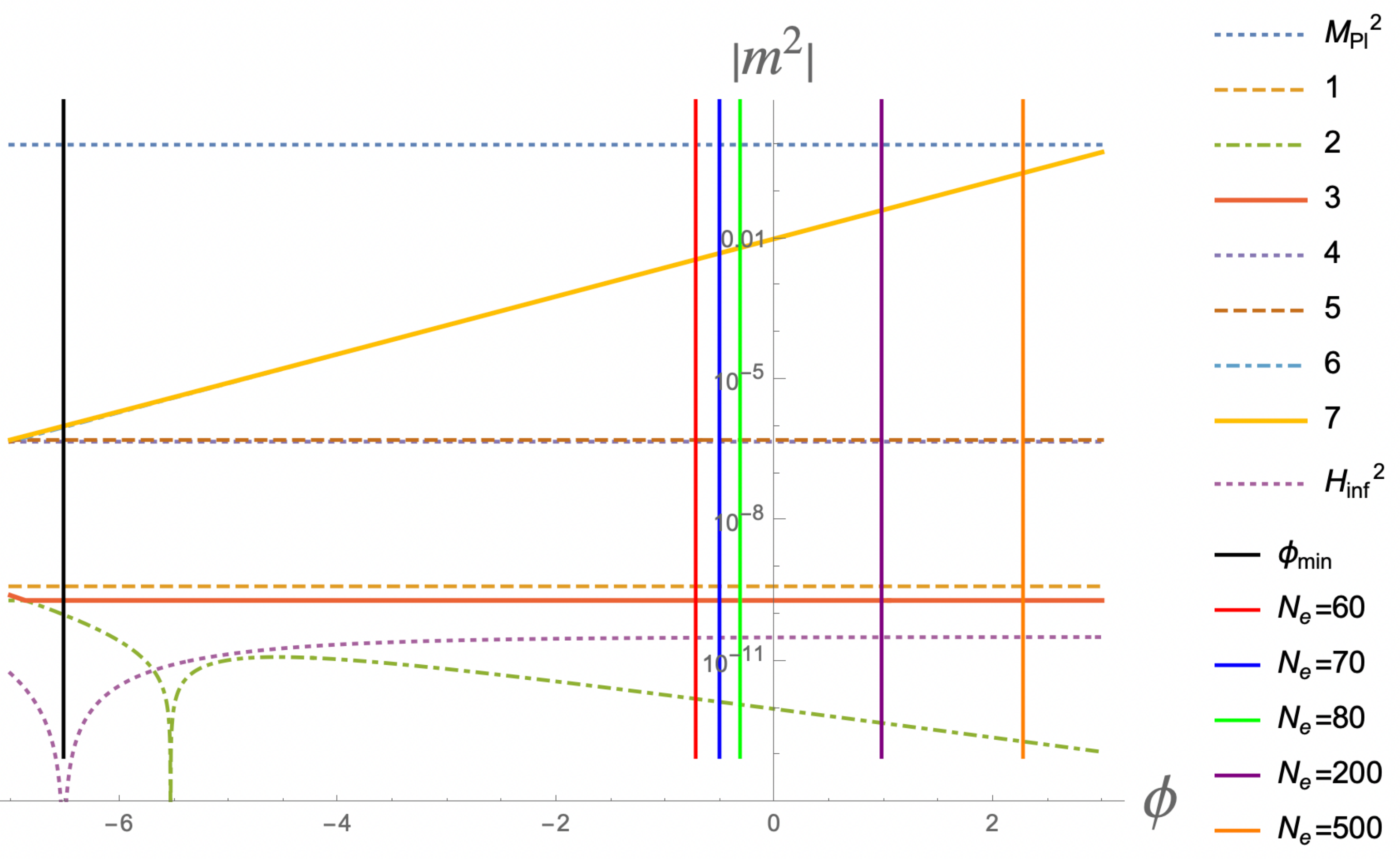}  \hskip 12pt
\includegraphics[scale=0.23]{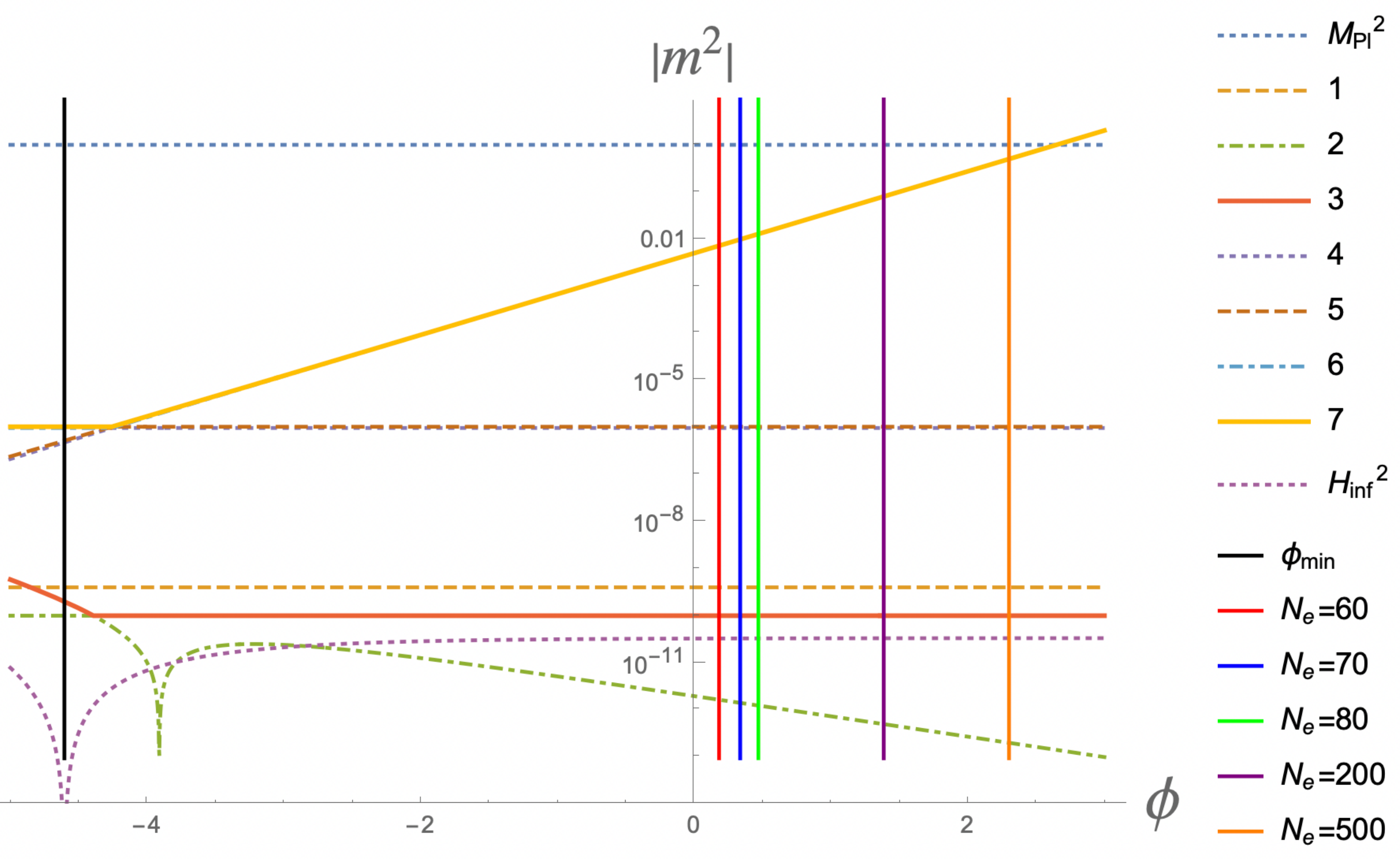}
\end{center}
\vspace{-.12cm}
\caption{\small
 On the left panel we present the case with $3\alpha=4$, on the right panel we present the case with $3\alpha=2$.}
\label{fig:split}
\end{figure}
\begin{figure}[!h]
\vspace{-1mm}
\hspace{-3mm}
\begin{center}
 \includegraphics[scale=0.23]{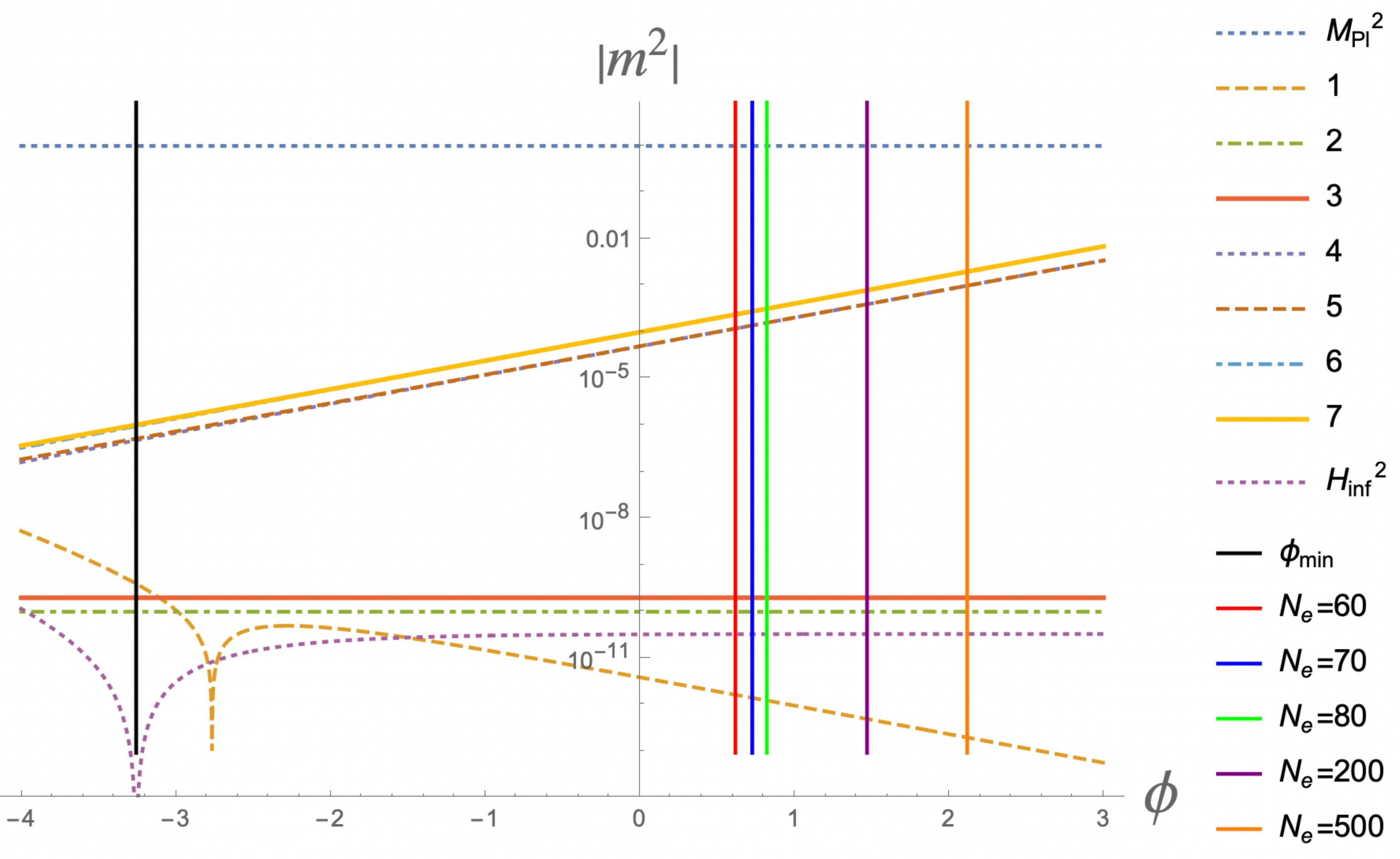}  \hskip 28pt
\includegraphics[scale=0.3]{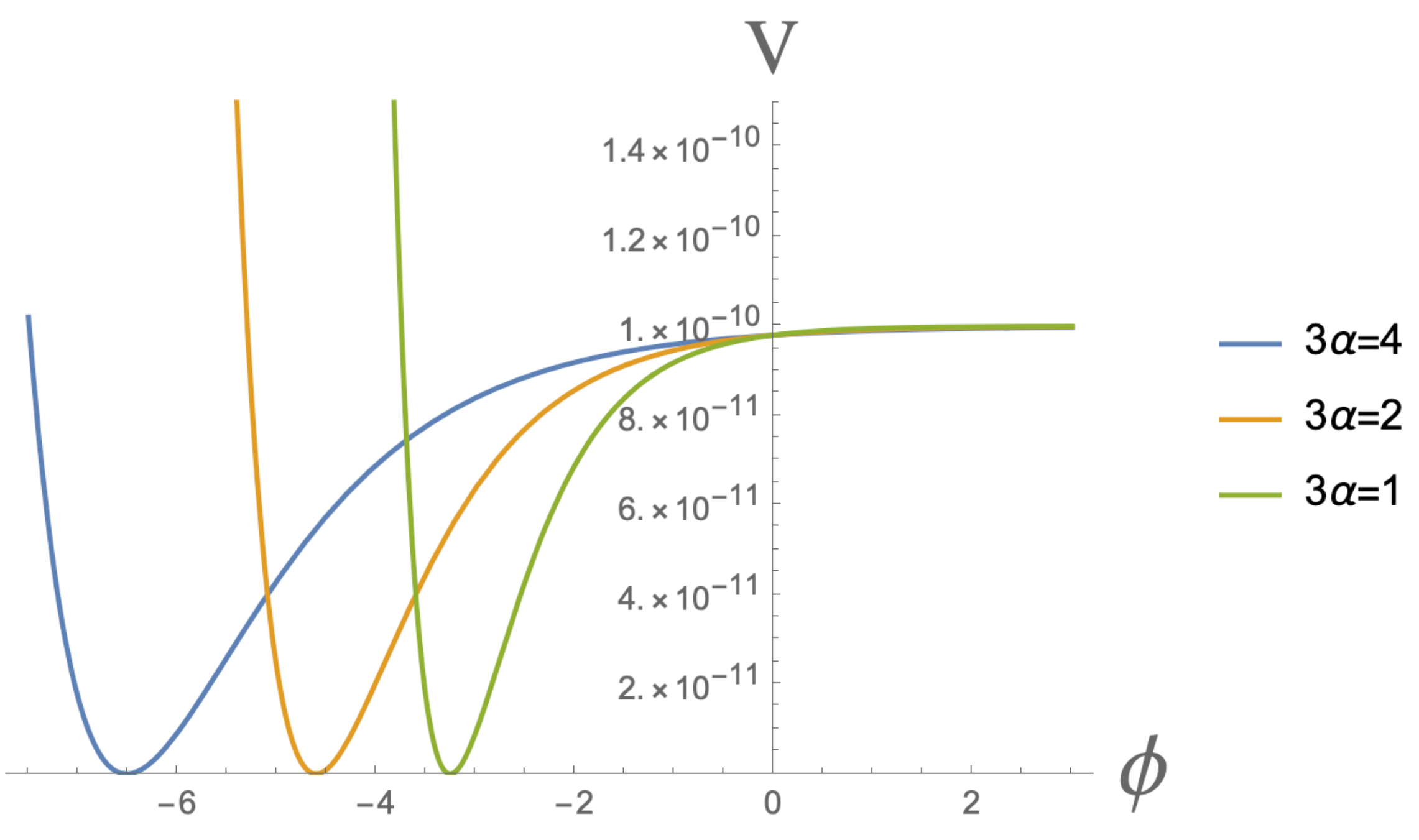}
\end{center}
\vspace{-.12cm}
\caption{\small
 On the left panel we present the case with $3\alpha=1$,  on the right panel we present the inflaton potentials for the cases with $3\alpha=4,2,1$.}
\label{fig:split1}
\end{figure}
For $c=10^2, W_0= 10^{-5}, m= 10^{-5}, \Lambda =0$ we present in Fig.~\ref{fig:split} the dynamics of scalars fields for the $3\alpha=4$ and $3\alpha=2$ models. In  Fig.~\ref{fig:split1} we show at the left panel the dynamics of scalars fields with $3\alpha=1$  and on the right panel we show the inflationary potentials for all these models.

To summarize, we now have derived $3\alpha=7, 4, 2, 1$ models  with cosmological predictions  shown in Fig. \ref{7disk2}.

\

\section{Merger of $\mathbf{(4,2,1)}$ split model into $\mathbf{(6,1)}$, $\mathbf{(5,2)}$, $\mathbf{(4,3)}$ }
We discussed the merger mechanism of split models with few flat directions into a model with less flat directions in \cite{Kallosh:2017ced, Kallosh:2017wnt, Gunaydin:2020ric}. 
We enforce the merger of the $(4,2,1)$ split model by adding to $V_{\rm infl} $ terms which require two different flat directions to coincide. For example, we can add one of the terms of the form
\be
M_1^2(T_{(1)}- T_{(2)}) (\overline T_{(1)}-\overline  T_{(2)}),  \quad M_2^2(T_{(1)}- T_{(3)}) (\overline  T_{(1)}- \overline T_{(3)}),  \quad  M_3^2(T_{(2)}- T_{(3)}) (\overline  T_{(2)}- \overline T_{(3)})
\ee
This will result in models with two flat directions, $(6,1)$, $(5,2)$, $(4,3)$, respectively. We have checked that merger parameters of the order $M_i\sim 10^{-3}$ are working well for our purpose. Namely, the heavy fields remain heavy and there are 2 light fields representing a two valley cosmological scenario. This one, as we know from earlier studies, for example in 
\cite{Gunaydin:2020ric}, allows us to obtain E-models with all discrete values of $3\alpha =7,6,5,4,3,2,1$.

The related T-models where the scalar potentials are instead of $\Big (1- e^{\sqrt{2\over 3\alpha} \phi}\Big)^2$ given by \\
$\tanh^{2}
\Big (\frac{1}{\sqrt{6\alpha}} \phi\Big)$. The relation between these models was explained in Sec.~\ref{ET}. We presented above in the introduction the predictions for both classes of models in Fig.~\ref{7disk2}.

\section{Discussion}

\begin{figure}[H]
\centering
\includegraphics[scale=0.16]{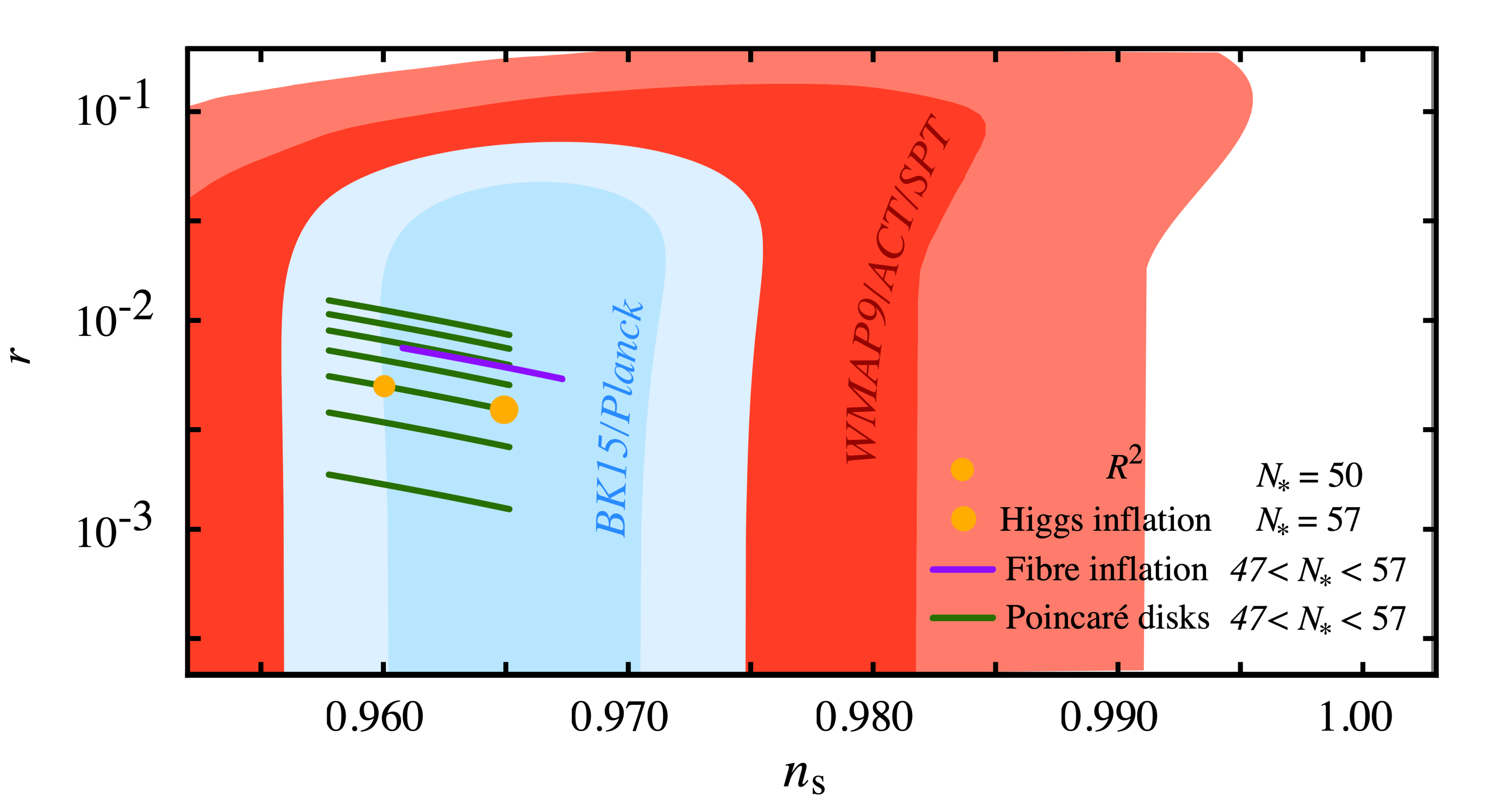}
\caption{\small The figure is courtesy of R. Flauger.  It shows the main targets for future detection of B-modes. The   seven  dark green lines in the $n_s$ - $r$ plane correspond to Poincar\'e disks models. We describe these models in eq. \rf{top71}. Each  of the 7 Poincar\'e disks in this figure corresponds to a potential  $V =V_0 Z\bar Z=  V_{0} \tanh^{2}({\phi/ \sqrt{6\alpha}})$ with  $3\alpha =1,2,3,4,5,6,7$.}
\label{Flauger}
\end{figure}

Plateau potentials, like $\alpha$-attractors which we show in  Fig.~\ref{Flauger} appear to be in a good agreement with the CMB data. The models of sequestered inflation studied in this paper easily accommodate plateau potentials. In fact, these models suggest a possible origin of plateau potentials.    These will be tested during the next 2 decades, but at present they appear to be good candidates of inflationary models which fit the data.

Sequestering means that we take an exactly flat direction in the fundamental theory, type IIB string theory in this paper,  at  Step I and uplift this flat direction to a plateau potential at Step II. The flat directions, one or three in our models in this paper, are in one to one correspondence to Nambu-Goldstone supermultiplets. 

According to the Nambu-Goldstone theorem for supermultiplets in supergravity with Minkowski vacua, see Appendix~\ref{app:Goldstone}, the flux superpotentials  must have certain symmetries, as many as the number of  the massless Goldstone multiplets. We have found these symmetries and presented our fluxes in Secs.~\ref{one} and \ref{three} in the form in which the symmetries are manifest. The key to this feature is that $W_{\rm flux}(T^i)$ with $i=1, \dots, 7$ in our models actually depends only on some differences between moduli, namely all new flux potentials are of the form $W_{\rm flux}(T^i-T^j)$.
We have encountered the analogous feature in M-theory flux superpotentials in \cite{Gunaydin:2020ric}.
The symmetries of these new flux superpotentials allows one to make an arbitrary holomorphic change of variables which preserve the differences $T^i-T^j$.

 As usual with Goldstone fields, only non-perturbative quantum corrections may uplift the mass and convert the flat direction into a nearly flat  inflationary plateau potential, as we have shown in  \cite{Gunaydin:2020ric} for M-theory and here in type IIB string theory.

To summarize, already at the Step I in M-theory compactified on twisted 7-tori with $G_2$-holonomy and in type IIB string theory compactified on a ${\mathbb{T}^6\over \mathbb{Z}_2 \times \mathbb{Z}_2}$ orientifold with generalized fluxes and O3/O7-planes we find flux superpotentials with flat direction(s) and hyperbolic geometries from the \K\,potentials. At Step II these models are naturally converted into plateau potentials which at present fit the data very well and therefore provide exciting targets for future observations.

Note also that the models with discrete values for the $3\alpha$-parameter are associated with string theory, M-theory and maximal supersymmetry. These are the seven targets shown in Fig.~\ref{Flauger}, at the specific seven values of the parameter $r$ which defines the B-modes. Meanwhile, for continuous values of the $3\alpha$-parameter there is a band of values of $r$ as we have shown in Fig.~\ref{7disk2}. It is possible that the B-modes will be discovered above or below the seven hyperbolic disks targets, and still fit the data on $n_s$. B-modes  may not even be detected if $r\ll 10^{-3}$.
If, however,
 the future data on B-modes will fit one of the seven discrete targets, the cosmological models  associated  with string theory, M-theory, maximal supergravity will get a strong support as  the favorite models of theoretical physics which fit the cosmological observations. 

\subsection{Addendum}

After our paper was submitted, BICEP/Keck released its latest constraints on the tensor/scalar ratio $r$, which are the most stringent constraints up to date \cite{BICEPKeck:2021gln}.  
We decided to add here Fig. 5 from \cite{BICEPKeck:2021gln} in combination with the predictions of the simplest T-models and E-models of $\alpha$-attractors. As we already discussed, these models have a general prediction for the tensor/scalar ratio $r = {12\alpha\over N_{e}^{2}}$, where $N_{e} \sim 50 - 60$ is the number of e-foldings during the last stages of inflation responsible for structure formation in the observable part of the universe. 

In general, the parameter $\alpha$ can take any value. Therefore $r$ also can take a broad range of values, all the way down to $r = 0$. However, in the models motivated by maximal supergravity, M-theory and string theory, which we studied in the present paper, there are 7 especially interesting values  $3\alpha =1,2,3,4,5,6,7$, see Figs. 1 and 7.  For the largest one in this series, $3\alpha = 7$, one has $r \approx 0.01$, which is very close to the range of $r$ explored in the recent BICEP/Keck paper \cite{BICEPKeck:2021gln}.
\begin{figure}[H]
\centering
\includegraphics[scale=0.45]{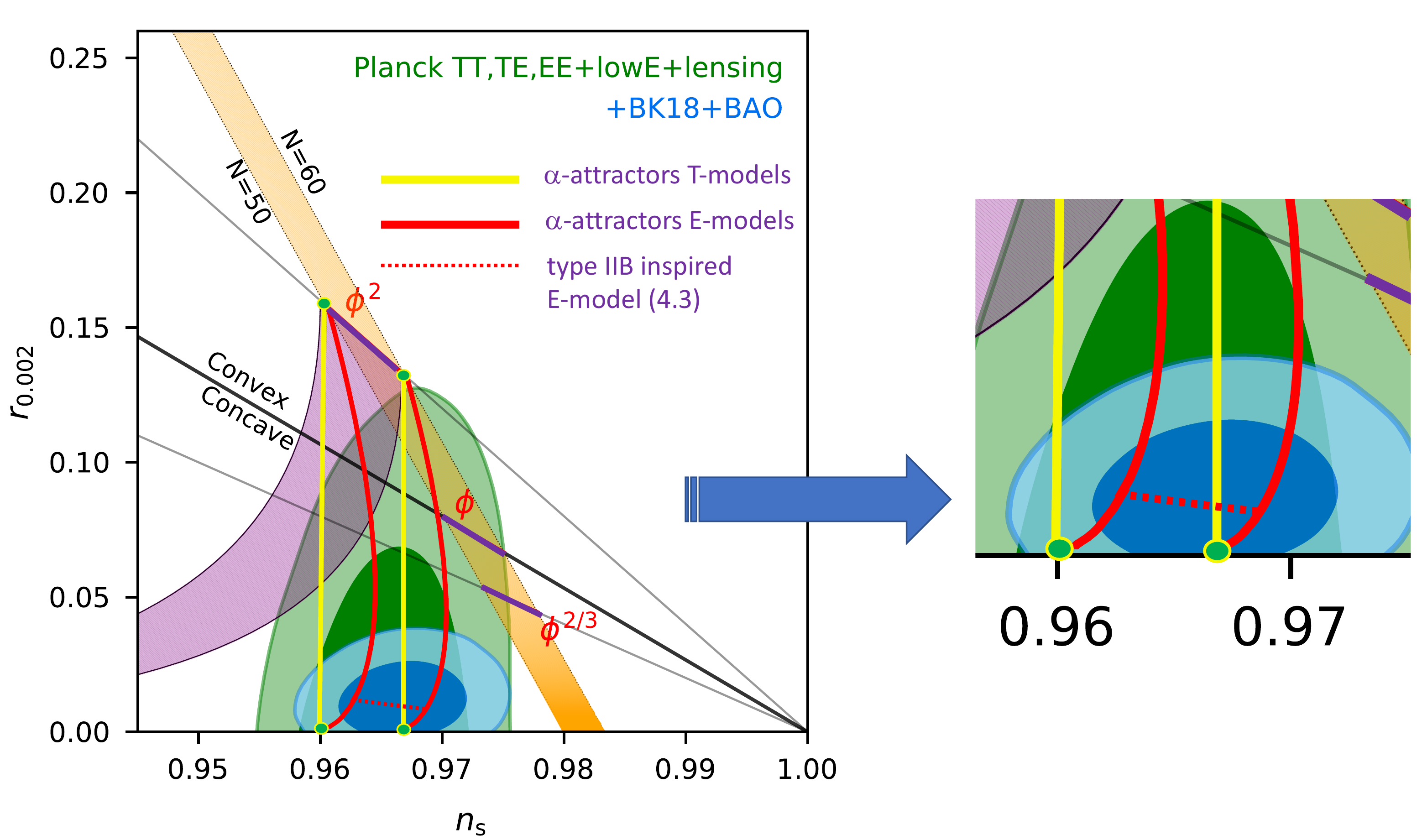}
\caption{\small The figure illustrating the main results of the BICEP/Keck2021 data release \cite{BICEPKeck:2021gln} superimposed with the predictions of the simplest $\alpha$-attractor models. The predictions of the type IIB string  theory  inspired scenario with the inflaton potential \rf{Echoice} - \rf{Epot} corresponding to $3\alpha = 7$ are shown by the red dashed line at the center of the dark blue area favored by BICEP/Keck.}
\label{BICEP}
\end{figure}

Predictions of the simplest T-models (E-models) are shown in Fig. \ref{BICEP} by two yellow (red) lines corresponding to $N_{e} = 50$ and $60$. The red dashed line represents predictions of the single valley type IIB scenario  \rf{Echoice} - \rf{Epot}. This dashed line is at the center of the dark blue area favored by the latest BICEP/Keck data \cite{BICEPKeck:2021gln}. At present the error bars of their result $r = 0.014 \pm 0.01$ are large, $\sigma(r) \sim 0.009$. However, the authors of  \cite{BICEPKeck:2021gln} expect that within the next few years they may improve the accuracy up to $\sigma(r) \sim 0.003$.

\ 

\section*{Acknowledgement}
We are grateful to  G. Dall'Agata, S. Ferrara,  R. Flauger,  C. L. Kuo, D. Roest and   A. Van Proeyen  for stimulating discussions.  RK and AL are supported by SITP and by the US National Science Foundation Grant  PHY-2014215, and by the  Simons Foundation Origins of the Universe program (Modern Inflationary Cosmology collaboration).  YY is  supported by JSPS KAKENHI, Grant-in-Aid for JSPS Fellows JP19J00494. TW is supported in part by the US National Science Foundation Grant  PHY-2013988.

\appendix

\section{Tadpole cancellation}\label{app:Tadpoles}
We consider  type IIB string theory compactified on the ${\mathbb{T}^6\over \mathbb{Z}_2 \times \mathbb{Z}_2}$ orientifold with O3/O7-planes and geometric and non-geometric fluxes. In the conventions of \cite{Aldazabal:2006up} the superpotential  $W$ can have terms of order $0,1,2,3,4,5$ in the fields and is given by
\begin{align}
W=&e_0+{\rm i}\sum_{I=1}^3e_IU_{I}-\sum_{I=1}^3 q_I\frac{U_1U_2U_3}{U_I}+{\rm i}mU_1U_2U_3\nonumber\\
&+S\left[{\rm i}h_0-\sum_{I=1}^3a_IU_I+\sum_{I=1}^3{\rm i} \bar{a}_I\frac{U_1U_2U_3}{U_I}-\bar{h}_0U_1U_2U_3\right]\nonumber\\
&+\sum_{I=1}^3T_I\left[-{\rm i}h_I-\sum_{J=1}^3U_Jb_{JI}+\sum_{J=1}^3{\rm i}\bar{b}_{JI}\frac{U_1U_2U_3}{U_J}+\bar{h}_IU_1U_2U_3\right]-S\sum_{I=1}^3f_IT_I\nonumber\\
&+\sum_{I,J=1}^3{\rm i}g_{JI}SU_JT_I+\sum_{I,J=1}^3\bar{g}_{JI}ST_I\frac{U_1U_2U_3}{U_J}-{\rm i}SU_1U_2U_3\sum_{I=1}^3\bar{f}_IT_I\,.
\label{general}\end{align}
The are a variety of tadpole cancellation conditions and Bianchi identities that the fluxes are required to satisfy. Some of them potentially require the presence of local sources like D3, D7, NS7 or I7 branes. Below is the full list with reference to their derivation in \cite{Aldazabal:2006up} (see also \cite{Guarino:2008ik}) .

 
\noindent D3 brane number: (2.35) in \cite{Aldazabal:2006up}
\begin{equation}
N_{\rm D3}=16-\frac12 \left[mh_0-e_0\bar{h}_0+\sum_{I=1}^3(q_Ia_I+e_I\bar{a}_I)\right].
\label{D3}\end{equation}

\noindent D7 brane number: (3.9) in \cite{Aldazabal:2006up} (after adding the contributions from O7-planes, see also equation (3.19) and the text below it in \cite{Guarino:2008ik})
\begin{equation}
N_{{\rm D7}_I}=16 +\frac12 \left[mh_I-e_0\bar{h}_I-\sum_{J=1}^3(q_Jb_{JI}+e_J\bar{b}_{JI})\right].
\end{equation}

\noindent NS7 brane number: (4.13) in \cite{Aldazabal:2006up}
\begin{equation}
N_{{\rm NS7}_I}=\frac12 \left[h_0\bar{f}_I-\bar{h}_0f_I-\sum_{J=1}^3(\bar{a}_Jg_{JI}-a_J\bar{g}_{JI})\right].
\end{equation}

\noindent I7 brane number: (4.40) in \cite{Aldazabal:2006up}
\begin{equation}
N_{{\rm I7}_I}=-\frac12 \left[e_0\bar{f}_I - m f_I + \sum_{J=1}^3(q_J g_{JI}+e_J\bar{g}_{JI})\right].
\end{equation}

\noindent $QH-PF=0$ constraints: (4.32)-(4.35) in \cite{Aldazabal:2006up} for $I\neq J \neq K \neq I$
\begin{align}
\bar{h}_0h_J+\bar{a}_Ib_{IJ}+\bar{a}_J\bar{b}_{JJ}-a_K\bar{b}_{KJ}+mf_J-q_Ig_{IJ}-q_Jg_{JJ}-e_K\bar{g}_{KJ}=0,&\\
h_0\bar{h}_J+a_I\bar{b}_{IJ}+a_{J}\bar{b}_{JJ}-\bar{a}_Kb_{KJ}-e_0\bar{f}_J-e_I\bar{g}_{IJ}-e_J\bar{g}_{JJ}-q_Kg_{KJ}=0,&\\
\bar{h}_0b_{KJ}+\bar{a}_I\bar{b}_{JJ}+\bar{a}_J\bar{b}_{IJ}-a_K\bar{h}_J+mg_{KJ}-q_I\bar{g}_{JJ}-q_J\bar{g}_{IJ}-e_K\bar{f}_J=0,&\\
h_0\bar{b}_{KJ}+a_Ib_{JJ}+a_Jb_{IJ}-\bar{a}_Kh_J-e_0\bar{g}_{KJ}-e_Ig_{JJ}-e_Jg_{IJ}-q_Kf_J=0.&
\end{align}

\noindent $QQ=0$ constraints: (3.30)-(3.33) in \cite{Aldazabal:2006up} for $I\neq J \neq K \neq I$
\begin{align}
-b_{II}b_{JK}+\bar{b}_{KI}h_K+h_I\bar{b}_{KK}-b_{JI}b_{IK}=0,&\\
-\bar{b}_{II}\bar{b}_{JK}+b_{KI}\bar{h}_K+\bar{h}_Ib_{KK}-\bar{b}_{JI}\bar{b}_{IK}=0,&\\
-b_{II}\bar{b}_{IJ}+\bar{b}_{JI}b_{JJ}+h_I\bar{h}_{J}-b_{KI}\bar{b}_{KJ}=0,&\\
\bar{b}_{II}b_{IJ}-b_{JI}\bar{b}_{JJ}+h_I\bar{h}_J-b_{KI}\bar{b}_{KJ}=0.&
\end{align}

\noindent $PP=0$ constraints: (4.16)-(4.19) in \cite{Aldazabal:2006up} for $I\neq J \neq K \neq I$
\begin{align}
-g_{II}g_{JK}+\bar{g}_{KI}f_K+f_I\bar{g}_{KK}-g_{JI}g_{IK}=0,&\\
-\bar{g}_{II}\bar{g}_{JK}+g_{KI}\bar{f}_K+\bar{f}_Ig_{KK}-\bar{g}_{JI}\bar{g}_{IK}=0,&\\
-g_{II}\bar{g}_{IJ}+\bar{g}_{JI}g_{JJ}+f_I\bar{f}_J-g_{KI}\bar{g}_{KJ}=0,&\\
\bar{g}_{II}g_{IJ}-g_{JI}\bar{g}_{JJ}+f_{I}\bar{f}_J-g_{KI}\bar{g}_{KJ}=0.&
\end{align}

\noindent $QP+PQ=0$ constraints: (4.24)-(4.27) in \cite{Aldazabal:2006up} for $I\neq J \neq K \neq I$
\begin{align}
b_{KK}\bar{g}_{KJ}-h_K\bar{f}_J-\bar{b}_{JK}g_{JJ}+b_{IK}\bar{g}_{IJ}+g_{KK}\bar{b}_{KJ}-f_K\bar{h}_{J}-\bar{g}_{JK}b_{JJ}+g_{IK}\bar{b}_{IJ}=0,&\\
b_{KK}g_{IJ}-h_K\bar{g}_{JJ}-\bar{b}_{JK}f_{J}+b_{IK}g_{KJ}+g_{KK}b_{IJ}-f_K\bar{b}_{JJ}-\bar{g}_{JK}h_J+g_{IK}b_{KJ}=0,&\\
\bar{b}_{KK}\bar{g}_{IJ}-\bar{h}_Kg_{JJ}-b_{JK}\bar{f}_{J}+\bar{b}_{IK}\bar{g}_{KJ}+\bar{g}_{KK}\bar{b}_{IJ}-\bar{f}_Kb_{JJ}-g_{JK}\bar{h}_{J}+\bar{g}_{IK}\bar{b}_{KJ}=0,&\\
\bar{b}_{KK}g_{KJ}-\bar{h}_Kf_J-b_{JK}\bar{g}_{JJ}+\bar{b}_{IK}g_{IJ}+\bar{g}_{KK}b_{KJ}-\bar{f}_Kh_J-g_{JK}\bar{b}_{JJ}+\bar{g}_{IK}b_{IJ}=0.&
\label{QP}\end{align}

\section{Symmetries of Superpotentials  and Nambu-Goldstone supermultiplets}\label{app:Goldstone}
The Nambu-Goldstone theorem  was proven in global supersymmetric theories in \cite{Kugo:1983ma}.  It was shown that for each massless supermultiplet there is a symmetry of the superpotential. We present here the generalization of this theorem to supergravity.
 
The  fermion spin 1/2 mass matrix in Minkowski vacua in supergravity with $W= W_{,i} =0$ is
\be
m_{ij} (T^k)\Big |_{W= W_{,i}=0} =e^{K\over 2} W_{,ij} (T^k), \qquad i=1,\dots, M.
\ee
The Minkowski vacuum is  at $T^k= T^k_0$
\be
W\Big | _{T^k_0} = {\delta W\over \delta T^i} \Big | _{T^k_0}=0.
\label{Mink}\ee
The Goldstone theorem for  chiral multiplets $T^i$ can be formulated as follows.
For each zero eigenvalue of $m_{ij}$ in Minkowski vacuum, in case there are $L$ of them,  there is a spontaneously broken symmetry of $W$. Namely, one should be able to establish a symmetry of $W$ under the following continuous transformations
\be
T^i \Rightarrow T^i + \alpha^a \Delta ^{i a} (T^l), \qquad a=1,\dots, N.
\label{sym2}\ee 
Proof:  Assume $W$ has a set of $N$ symmetries 
\be
{\delta W\over \delta T^i}  \Delta ^{i a} (T^l) =0.
\label{Wsym}\ee
We differentiate this equation over $T^j$ and view it at $W= W_{,i}=0$, i. e. at $T^k= T^k_0$.
Taking into account eq. \rf{Mink} we find
\be
\Big [ {\delta^2 W\over \delta T^i\delta T^l} \Delta ^{i a}(T^l)\Big ]_{T^i = T^i_0}=0.
\label{Wbr_sym}\ee
The symmetry generators $\Delta ^{i a}(T^i)$ at $T^i = T^i_0$ either vanish or not
\bea\label{SB}
\Delta_{ SB} ^{i a}(T^i_0) &&\neq 0\, , \qquad  a=1,\dots, L  \\
\cr
 \Delta_{UB} ^{i a}(T^i_0) &&= 0 ,\qquad  a=L+1,\dots, N.
\eea
Under the transformation \rf{sym2} the ground state transforms as follows
\bea
T^i_0 &\rightarrow& T^i_0 + \alpha^a 
\Delta_{SB} ^{i a}(T^i_0), \qquad a=1,\dots, L\,\\
T^i_0 &\rightarrow& T^i_0 \ , \qquad \qquad \qquad\qquad\! a= L+1,\dots, N.
\eea
Therefore the $L$ symmetries which affect the ground state, are qualified as spontaneously broken symmetries. The remaining $N-L$ symmetries remain symmetries of the ground state, they are qualified as  unbroken symmetries.\footnote{We show the possible unbroken generators here to keep generality of the proof. However, our examples only contain spontaneously broken (nonlinearly realized) symmetry generators.} Equation \rf{Wbr_sym}  acquires the form
\be
\Big [ {\delta^2 W\over \delta T^i\delta T^l} \Delta_{SB} ^{i a}(T^l)\Big ]_{T^i = T^i_0}=0.
\label{mass}
\ee
It shows that the rank of the $M\times M$ mass matrix $W_{,ij} (T^k_0)$ is less than $M$, and it is less or equal to $M-L$, since there are $L$  non-vanishing eigenvectors. It means that for each of the symmetries in eq. \rf{SB} we are bound to find a massless chiral multiplet. And vice versa, 
for each of the $L$ flat directions of $W$ in Minkowski minimum we are bound  to find the corresponding $L$ symmetries of $W$.

{ \it Examples}

\noindent Our first example involves the octonion superpotentials of the M-theory in  \cite{Gunaydin:2020ric} where
\be
W_{\rm oct}= \sum_{\{ijkl \}} (T^i -T^j) (T^k -T^l),\qquad i=1,\dots , 7
\label{ourW} \ee
\noindent where we  take a sum over  7 different 4-qubit states defining the choice of ${\{ijkl \} }$ in $W_{\rm oct}$. For example we can take
    \bea\label{cw}
&& W_{\rm oct} = (T^2 - T^4) (T^5 -T^6)+  (T^3 - T^5) (T^6 - T^7) + (T^4 - T^6) (T^7 -
      T^1)\\
 \cr
      && + (T^5 - T^7) (T^1 - T^2) + (T^6 - T^1) (T^2 - T^3) + (T^7 -
      T^2) (T^3 -T^4) + (T^1 - T^3) (T^4 - T^5).  \nonumber
   \eea 
   One important property of the superpotentials  $W_{\rm oct}$ is the fact that under an arbitrary holomorphic shift all fields are shifted by the same holomorphic function
\be
T^i \Rightarrow T^i + F(T^k).
\ee
The difference between two fields and therefore $W_{\rm oct}$ are invariant 
\be
(T^i -T^j) \Rightarrow (T^i -T^j),  \qquad W_{\rm oct}  \Rightarrow W_{\rm oct} .
\ee
The ground state solution of the equations for the Minkowski vacuum is  at $T^k= T^k_0$. For a given choice of $T^k_0$ the symmetry of  $W_{\rm oct} $ is broken. According to the theorem above, one of the eigenvalues of the mass matrix of the supermultiplets must vanish. This is indeed the case.

The second example involves the M-theory superpotential with two flat directions
\begin{eqnarray}
W_{ ( 5.2)} = && (T^5 -T^7) (T^1 -T^2)+ (T^6 -T^1) (T^2 -T^3)+(T^7 -T^2) (T^3 -T^4)\cr
\cr
&& + (T^1 -T^3) (T^4 -T^5)+ (T^2 -T^4) (T^5 -T^6).
\label{52}\end{eqnarray}
The flat directions are $T^5=T^6$ and $T^1=T^2=T^3 = T^4= T^7 $. One finds that there are two  symmetries of $W_{ ( 5.2)} $ are
\be 
 T^5\rightarrow T^5 + F_1(T^k) \ , \qquad
 T^6=T^6 + F_1(T^k) \ , 
\ee 
and 
\bea
&&T^1\rightarrow T^1 + F_2(T^k) \ , \qquad T^2=T^2 + F_2(T^k)\ ,\qquad T^3=T^3 + F_2(T^k)\ , \cr
&&T^4\rightarrow T^4 + F_2(T^k) \ , \qquad
 T^7=T^7 + F_2(T^k) \ .
\eea

All these symmetries are broken on a ground state which is at $T^k= T^k_0$. The conditions of the theorem are satisfied and there are two massless states in this model. 

Examples of Nambu-Goldstone supermultiplets in type IIB string theory are given in Secs.~\ref{one} and \ref{three}.

\newpage

\bibliographystyle{JHEP}
\bibliography{lindekalloshrefs}
\end{document}